\documentclass[journal,draftclsnofoot,onecolumn]{IEEEtran}

\newtheorem{theorem}{Theorem}
\newtheorem{corollary}{Corollary}

\newtheorem{lemma}{Lemma}
\newtheorem{definition}{Definition}
\newtheorem{example}{Example}
\newtheorem{remark}{Remark}

\newcommand{\comment}[1]{}

\usepackage{graphicx}
\usepackage{amssymb}
\usepackage{amsmath}
\usepackage{eufrak}
\usepackage{float}
\usepackage{hyperref}
\usepackage{wasysym}
\usepackage{marvosym}
\usepackage{ifsym}
 \usepackage{stmaryrd} 
\usepackage{xcolor}
\begin{document}
%
\title{Message Transmission over Classical Quantum
Channels     with a Jammer with Side Information: Message Transmission
Capacity and Resources   }
\author{Holger Boche\allowdisplaybreaks\\
Lehrstuhl f\"ur Theoretische\allowdisplaybreaks\\
Informationstechnik,\allowdisplaybreaks\\ Technische Universit\"at  M\"unchen,\allowdisplaybreaks\\
 Munich, Germany,\allowdisplaybreaks\\
  boche@tum.de \allowdisplaybreaks\\
Minglai Cai\allowdisplaybreaks\\
Lehrstuhl f\"ur Theoretische\allowdisplaybreaks\\                                                                                                                                                                                           
Informationstechnik,\allowdisplaybreaks\\ Technische Universit\"at  M\"unchen,\allowdisplaybreaks\\
Munich, Germany,\allowdisplaybreaks\\                                                                                                                                                                                                                      
  minglai.cai@tum.de\allowdisplaybreaks\\
Ning Cai\allowdisplaybreaks\\
School of Information Science\allowdisplaybreaks\\
 and Technology,\allowdisplaybreaks\\ 
ShanghaiTech University,\allowdisplaybreaks\\
Shanghai, China,\allowdisplaybreaks\\
ningcai@shanghaitech.edu.cn }

%
%



\maketitle

\begin{abstract}
In this paper we
propose a new model for arbitrarily varying classical-quantum
  channels.
In this model a jammer  
has side information.
We consider two scenarios.
In the first scenario
the jammer
knows the channel input,
while in the second  scenario
the jammer
knows both the
channel input and the message.
The transmitter and receiver share a secret random
key with a vanishing
key rate. We determine the capacity for
both average and maximum error 
criteria     for both scenarios.
We also establish the strong converse. 
We show
that all these corresponding capacities
are equal,
which means      that additionally
revealing the message to the jammer does not change the capacity.

\end{abstract}


%
\IEEEpeerreviewmaketitle

\section{Introduction} 
Quantum information theory has developed into a very active
field of reseach in the last years and its study provide an enormous amount of potential advantages. 
Quantum channels differs significantly from communication over
classical channels.
Quantum communication allow us to exploit  
possibilities
for new applications for communications. To name a few: 
message transmission, secret message transmission, entanglement transmission,
entanglement generation.   secure communications over quantum channels
is one of  the first practical applications of quantum communications.
In such systems one usually consider active jamming and passive
eavesdropping attacks.

  Communication 
models including a jammer who tries to disturb the legal parties' 
 communication  have received a  lot of attention in recent years.
These publications concentrated on the model of message transmission over
an arbitrarily varying channel
where a third channel user, the jammer,  may change his input in every channel use.
This model captures completely all possible jamming attacks
 and is not restricted to use a 
repetitive probabilistic strategy.
The arbitrarily varying channel was  introduced
 in \cite{Bl/Br/Th2}.  In the model of message transmission over arbitrarily varying 
 channels
it is understood that the sender and the receiver have to select
their coding scheme first. 
In the conventional model 
it is assumed that this coding scheme is known by the
jammer, and he may choose the most advantaged jamming attacking strategy
depending on his knowledge,
but the jammer has neither knowledge about the transmitted codeword
nor  knowledge about the message.
Ahlswede showed in \cite{Ahl1} the surprising  result,
  that either the deterministic capacity of an
arbitrarily varying channel is zero or it is equal to its random correlated
 capacity (Ahlswede dichotomy).      For this dichotomy it is essential that
the average error criterion was used.     After 
that discovery, it remained an open question exactly 
when the deterministic capacity is nonzero. In \cite{Rei} Ericson gave a sufficient 
condition for that, and in \cite{Cs/Na} Csisz\'ar and Narayan proved that this is condition is also necessary.
Ahlswede dichotomy  demonstrates the importance of 
resources (shared randomness) in a very clear form.
It is required that both sender and receiver have access to a
perfect copy of the outcome of a random experiment,
and thus we should assume an additional perfect channel.
The legal channel users' knowledge about the shared randomness
is very helpful  for message transmission 
through an arbitrarily varying
channel (random correlated capacity), where  we assume that the resource is only known by 
the legal channel users, since otherwise it will be completely useless
(cf. \cite{Bo/Ca/De3}).

 In this work we consider classical quantum channels, i.e.,
the sender's  inputs are classical    symbols     and the receiver's outputs are quantum
systems. The  capacity
of classical-quantum channels     under
average error criterion       has been  determined  in
\cite{Ho}, \cite{Sch/Ni},
and  \cite{Sch/Wes}.
The capacity of arbitrarily varying classical-quantum
 channels has been delivered in \cite{Ahl/Bli}. 
An alternative proof of \cite{Ahl/Bli}'s result
and a proof of the
strong converse have been given in \cite{Bj/Bo/Ja/No}.  In \cite{Ahl/Bj/Bo/No}
Ahlswede dichotomy for the  arbitrarily varying classical-quantum
 channels was established, and a  sufficient and  necessary  condition
for  the  zero deterministic capacity has been given. In \cite{Bo/No} a simplification of this 
condition  was delivered.
See also \cite{Ka/Ma/Wi/Ya} and \cite{Ka/Ma/Wi/Ya2} for a classical quantum channel model
with a benevolent third channel user instead of with a jammer.
 These results are  basis tools for
 secure communication over 
   arbitrarily varying wiretap
 channels.
 An arbitrarily varying wiretap channel is a channel
with both a jammer and an eavesdropper. Classical arbitrarily varying wiretap channels
have been studied extensively in the
context of classical information theory.
The   secrecy capacity of  arbitrarily varying wiretap classical quantum channels has 
been determined in \cite{Bo/Ca/De3}.

As already mentioned        the message transmission capacity of
an arbitrarily varying channel depends on the demanded
error criterion. The deterministic
capacities of classical arbitrarily varying channel under
maximal error criterion and under
the average error criterion are in general, not equal.
The deterministic capacity formula of classical arbitrarily varying channels under
average error criterion is already well studied in the
context of classical information theory.
The deterministic capacity formula of classical arbitrarily varying channels under
maximal error criterion is still an open problem.
It 
has been shown by Ahlswede in \cite{Ahl0}  that the capacity under maximal error criterion of certain arbitrarily varying channels can be equal to the 
zero-error capacity of related discrete memoryless channels.
Furthermore
the random correlated capacities of  arbitrarily varying quantum to quantum channels under
maximal error criterion and under the average error criterion are equal.
Interestingly,  \cite{Bo/No}
shows that 
the deterministic capacities of  arbitrarily varying quantum to quantum channels under
maximal error criterion and under the average error criterion are equal,
since randomness      for encoding         is available for quantum to quantum channels,      i.e., quantum
encoding is very powerful.       By the above facts
there is no Ahlswede dichotomy for 
 arbitrarily varying channels under
maximal error criterion: It may occur that the 
 deterministic capacity  of a classical arbitrarily varying channel under
maximal error criterion is not zero, but on the other hand, unequal to
its  random correlated capacity.    We will provide a example in Section \ref{MR}.

In all the above mentioned works 
it is assumed that the jammer knows the coding scheme,
but has neither side information about the codeword
nor  side information about the  message of the legal transmitters.
In many applications,
especially for secure communications, it is too optimistic
to assume this.
Thus in this paper we want to consider two scenarios,
where the jammer has side information: 
In the first one
the jammer  knows both coding scheme and  input
codeword. 
In the second one
the jammer  knows additionally the message (cf. Figure \ref{scenario1}
and \ref{scenario2}).  The jammer can make use of this knowledge in
each scenario to
advance his attacking strategy. 
We
 require that 
information transmission can be guaranteed even in the worst case, when the jammer 
 chooses the most advantageous attacking strategy
according to his knowledge. For classical
arbitrarily varying 
 channels
this was first considered by \cite{Sar}.
In this paper we extend this  result to 
arbitrarily varying classical-quantum
 channels, where we use   techniques
different to these used in 
\cite{Sar} (cf. Section \ref{sec_prd}).
In this work we consider for  both scenarios
the random correlated
capacities under  average and maximal 
error criteria.
Detailed descriptions for both scenarios are
given in Section \ref{PF}. 
In Section \ref{MR} the message transmission
capacities for both scenarios and both error criteria
are completely
characterized. 
In Section \ref{sec_prd}, Section \ref{itswpt}, and
Section \ref{sec_prda} we
deliver proofs for the
capacities results for both scenarios and  both error criteria.
 A vanishing rate of the key is sufficient for our codes since
the resource we use here
is only of polynomial size of the code length
(cf. Remark \ref{iptiac}, and also \cite{Bo/No} and \cite{Bo/Ca/De2}
for a discussion about the difference between various forms of 
shared randomness).

\section{Problem Formulation}

\label{PF}
\textbf{\textit{A: Basic notations}}

Throughout the paper random variables will be  denoted by capital
letters e. g., $S,X,Y,$ and their realizations (or values) and
domains (or alphabets) will be denoted by corresponding lower case letters  e. g.,
$s,x,y,$ and script letters e.g., ${\cal S},{\cal X},{\cal Y}$, respectively. Random 
sequences will be denoted a by capital bold-face letters, whose lengths are understood by the context, e. g., 
${\bf S}=(S_1, S_2, \ldots, S_n)$ and ${\bf X}=(X_1,X_2, \ldots, X_n)$, and deterministic sequences 
are written as lower case bold-face letters e. g., ${\bf s}=(s_1,s_2, \ldots, s_n), {\bf x}=(x_1, x_2, \ldots, x_n)$. 

$P_X$ is distribution of random variable $X$. Joint distributions and conditional distributions of random variables $X$ and
 $S$ will be written as $P_{SX}$, etc  and $P_{S|X}$ etc, respectively and $P_{XS}^n$ and $P_{S|X}^n$ are their product distributions i. e.,
 $P_{XS}^n({\bf x},{\bf s}):= \prod_{t=1}^nP_{XS}(x_t,s_t)$, and 
$P_{S|X}^n({\bf s}|{\bf x}):=\prod_{t=1}^nP_{S|X}(s_t|x_t)$. 
Moreover ${\cal T}^n_X, {\cal T}^n_{XS}$ and ${\cal T}^n_{S|X}({\bf x})$ are sets 
of (strongly) typical sequences of the type $P_X$, joint type $P_{XS}$ and conditional type $P_{S|X}$, respectively. 
The cardinality of a set ${\cal X}$ will be denoted by $|{\cal X}|$. For a positive integer $L$, $[L]:=\{1,2, \ldots, L\}$. 
``$Q$ is a classical channel, or a conditional probability distribution, from set ${\cal X}$ to set ${\cal Y}$" is abbreviated to 
``$Q:{\cal X} \rightarrow {\cal Y}$". ``Random variables $X, Y$ and $Z$ form a Markov chain" is abbreviated to ``$X \leftrightarrow Y \leftrightarrow Z$". $\mathbb{E}$ will standard for the operator of mathematical expectation.

Throughout the paper dimensions of all Hilbert spaces are finite, 
and the identity operator in a Hilbert space ${\cal H}$ 
is denoted by $\mathbb{I}_{\cal H}$.

Throughout the paper the base(s) of logarithm  is  2.      For 
a discrete random variable $X$  on a finite set $\mathbf{X}$ and a discrete
random variable  $Y$  on  a finite set   $\mathbf{Y}$,   we denote the Shannon entropy
of $X$ by
$H(X)=-\sum_{x \in \mathcal{X}}p_x(x)\log p_x(x)$ and the mutual information between $X$
and $Y$ by  
$I(X;Y) = \sum_{x \in \mathcal{X}}\sum_{y \in \mathcal{Y}}  p_{x,y}(x,y) \log{ \left(\frac{p_{x,y}(x,y)}{p_x(x)p_y(y)} \right) }$.
Here $p_{x,y}$ is the joint probability distribution function of $X$ and $Y$, and 
$p_x$ and $p_y$ are the marginal probability distribution functions of $X$ and $Y$ respectively.

Let $\mathfrak{P}$ and $\mathfrak{Q}$ be quantum systems. We 
denote the Hilbert space of $\mathfrak{P}$ and $\mathfrak{Q}$ by 
$G^\mathfrak{P}$ and $G^\mathfrak{Q}$, respectively. Let $\phi^\mathfrak{PQ}$ be a bipartite
quantum state in $\mathcal{S}(G^\mathfrak{PQ})$. 
We denote the partial
trace over $G^\mathfrak{P}$ by
\[\mathrm{tr}_{\mathfrak{P}}(\phi^\mathfrak{PQ}):= 
\sum_{l} \langle l|_{\mathfrak{P}} \phi^\mathfrak{PQ} |  l \rangle_{\mathfrak{P}}\text{ ,}\]
where $\{ |  l \rangle_{\mathfrak{P}}: l\}$ is an orthonormal basis
of $G^\mathfrak{P}$.
We denote the conditional entropy by
\[S(\mathfrak{P}\mid\mathfrak{Q})_{\rho}:=
S(\phi^\mathfrak{PQ})-S(\phi^\mathfrak{Q})\text{
.}\]
Here $\phi^\mathfrak{Q}=\mathrm{tr}_{\mathfrak{P}}(\phi^\mathfrak{PQ})$.
\vspace{0.2cm}

 For a finite-dimensional
complex Hilbert space  ${\cal H}$, we denote
the (convex) set 
of  density operators on ${\cal H}$ by
\[\mathcal{S}({\cal H}):= \{\rho \in \mathcal{L}({\cal H}) :\rho  \text{ is Hermitian, } \rho \geq 0_{{\cal H}} \text{ , }  \mathrm{tr}(\rho) = 1 \}\text{ ,}\]
where $\mathcal{L}({\cal H})$ is the set  of linear  operators on ${\cal H}$, and $0_{{\cal H}}$ is the null
matrix on ${\cal H}$. Note that any operator in $\mathcal{S}({\cal H})$ is bounded.

 For  finite-dimensional
complex Hilbert spaces  ${\cal H}$ and  ${\cal H}'$  a \bf quantum channel \it $N$:
$\mathcal{S}({\cal H}) \rightarrow \mathcal{S}({\cal H}')$, $\mathcal{S}({\cal H})  \ni
\rho \rightarrow N(\rho) \in \mathcal{S}({\cal H}')$ is represented by a
completely positive trace-preserving map
 which accepts input quantum states in $\mathcal{S}({\cal H})$ and produces output quantum
states in  $\mathcal{S}({\cal H}')$. 

\textbf{\textit{B: Code definitions}}

If the sender wants to transmit a classical message of a finite set ${\cal X}$ to
the receiver using a quantum channel $N$, his encoding procedure will
include a classical-to-quantum encoder 
to prepare a quantum message state $\rho \in
\mathcal{S}({\cal H})$ suitable as an input for the channel. If the sender's
encoding is restricted to transmit an  indexed finite set of
 quantum states $\{\rho_{x}: x\in \mathcal{X}\}\subset
\mathcal{S}({\cal H})$, then we can consider the choice of the signal
quantum states $\rho_{x}$ as a component of the channel. Thus, we
obtain a channel $\sigma_x := N(\rho_{x})$ with classical inputs $x\in \mathcal{X}$ and quantum outputs,
 which we call a classical-quantum
channel. This is a map $\mathbf{N}$: $\mathcal{X} \rightarrow
\mathcal{S}({\cal H}')$, $\mathcal{X} \ni x \rightarrow \mathcal{N}(x) \in
\mathcal{S}({\cal H}')$ which is represented by the set of $|\mathcal{X}|$ possible
output quantum states $\left\{\sigma_x = \mathbf{N}(x) :=
N(\rho_{x}): x\in \mathcal{X}\right\}\subset \mathcal{S}({\cal H}')$, meaning that
each classical input of $x\in \mathcal{X}$ leads to a distinct quantum output
$\sigma_x \in \mathcal{S}({\cal H}')$. In view of this, we have the following
definition.\vspace{0.15cm}

\begin{definition}
Let ${\cal H}$ be a finite-dimensional
complex Hilbert space.
 A \bf classical-quantum channel   \it   is
a mapping $W: \mathcal{X}\rightarrow\mathcal{S}({\cal H})$,
 specified by a set of quantum states $\{\rho(x), x \in {\cal X}\}$  $\subset\mathcal{S}({\cal H})$, 
indexed by ``input letters" $x$ in a finite set ${\cal X}$. ${\cal X}$ and ${\cal H}$ 
are called input alphabet and output space respectively. 
We define the $n$-th extension of
 classical-quantum channel $W$ as follows.
The channel outputs a quantum state
$\rho^{\otimes n}({\bf x}):=\rho(x_1) \otimes \rho(x_2) \otimes \ldots, \otimes \rho(x_n)$,
in the $n$th tensor power ${\cal H}^{\otimes n}$ of the output space ${\cal H}$, when an 
input codeword ${\bf x}=(x_1,x_2, \ldots, x_n) \in {\cal X}^n$ of length $n$ is input into the channel.
\end{definition}

Let
$\mathtt{V}$: $\mathcal{X} \rightarrow
\mathcal{S}({\cal H})$ be a classical-quantum
channel.   For $P\in P(\mathcal{X})$,  
the conditional entropy of the channel for $\mathtt{V}$ with input distribution $P$
is denoted by
 \[S(\mathtt{V}|P) := \sum_{x\in {\cal X}} P(x)S(\mathtt{V}(x))\text{
.}\]

Let $\Phi := \{\rho_x : x\in \mathcal{X}\}$  be a
be a classical-quantum
channel, i.e., a
 set of quantum  states
labeled by elements of $\mathcal{X}$. For a probability distribution  $Q$
on $\mathcal{X}$, the    Holevo $\chi$ quantity is defined as
\[\chi(Q;\Phi):= S\left(\sum_{x\in \mathbf{A}} Q(x)\rho_x\right)-
\sum_{x\in \mathbf{A}} Q(x)S\left(\rho_x\right)\text{ .}\]

For a probability distribution $P$ on a finite set $\mathcal{X}$  and a positive constant $\delta$,
we denote the set of typical sequences by 
\[\mathcal{T}^n_{P,\delta} :=\left\{ x^n \in \mathcal{X}^n: \left\vert \frac{1}{n} N(x'\mid x^n)
- P(x') \right\vert \leq \frac{\delta}{|\mathcal{X}|}\forall x'\in \mathcal{X}\right\}\text{ ,}\]
where $N(x'\mid x^n)$ is the number of occurrences of the symbol $x'$ in the sequence $x^n$.\vspace{0.2cm}

Let
${\cal H}$ be a finite-dimensional
complex Hilbert space.
Let $n \in \mathbb{N}$ and $\alpha > 0$.
We suppose $\rho \in \mathcal{S}({\cal H})$ has
the spectral decomposition
$\rho = \sum_{x} P(x)  |x\rangle\langle x|$,
its
$\alpha$-typical subspace is the subspace spanned
by $\left\{|x^n\rangle, x^n \in {\mathcal{T}}^n_{P, \alpha}\right\}$,
where  $|x^n\rangle:=\otimes_{i=1}^n|x_i\rangle$.  The orthogonal subspace projector  which projected onto the
typical subspace is
\[ \Pi_{\rho ,\alpha}=\sum_{x^n \in {\mathcal{T}}^n_{P, \alpha}}|x^n\rangle\langle x^n|\text{ .}\]

Similarly,   let $\mathcal{X}$ be a finite set, and  $G$ be a finite-dimensional
complex Hilbert space.
Let
$\mathtt{V}$: $\mathcal{X} \rightarrow
\mathcal{S}({\cal H})$ be a classical-quantum
channel.   For $x\in\mathcal{X}$,   suppose
$\mathtt{V}(x)$ has
the spectral decomposition $\mathtt{V}(x)$
$ =$ $\sum_{j}
V(j|x) |j\rangle\langle j|$
for a stochastic matrix
$V(\cdot|\cdot)$.
 The $\alpha$-conditional typical
subspace of $\mathtt{V}$ for a typical sequence   $x^n$ is the
subspace spanned by
 $\left\{\bigotimes_{x\in\mathcal{X}}|j^{\mathtt{I}_x}\rangle, j^{\mathtt{I}_x} \in \mathcal{T}^{\mathtt{I}_x}_{V(\cdot|x),\delta}\right\}$.
Here $\mathtt{I}_x$ $:=$ $\{i\in\{1,\cdots,n\}: x_i = x\}$ is an indicator set that selects the indices $i$ in the sequence $x^n$
$=$ $(x_1,\cdots,x_n)$ for which the $i$-th
symbol $x_i$ is equal to $x\in\mathcal{X}$.
The subspace is often referred   to   as the $\alpha$-conditional typical
subspace of the state  $\mathtt{V}^{\otimes n}(x^n)$.
  The orthogonal subspace projector    which projected onto it   is defined as
	\[\Pi_{\mathtt{V}, \alpha}(x^n)=\bigotimes_{x\in\mathcal{X}}
\sum_{j^{\mathtt{I}_x} \in {\cal
T}^{\mathtt{I}_x}_{\mathtt{V}(\cdot \mid x^n),\alpha}}|j^{\mathtt{I}_x} \rangle\langle j^{\mathtt{I}_x}|\text{ .}
\]
The typical subspace 
has following  properties:

For $\sigma \in \mathcal{S}({\cal H}^{\otimes n})$ and $\alpha > 0$ 
there are positive constants $\beta(\alpha)$, $\gamma(\alpha)$, 
and $\delta(\alpha)$, depending on $\alpha$ and tending to zero
when $\alpha\rightarrow 0$ such that
\begin{equation} \label{te1}
\mathrm{tr}\left({\sigma} \Pi_{\sigma ,\alpha}\right) > 1-2^{-n\beta(\alpha)}
\text{ ,}\end{equation}

\begin{equation} \label{te2}
2^{n(S(\sigma)-\delta(\alpha))}\le \mathrm{tr} \left(\Pi_{\sigma ,\alpha}\right)
\le 2^{n(S(\sigma)+\delta(\alpha))}
\text{ ,}\end{equation}

\begin{equation} \label{te3}
2^{-n(S(\sigma)+\gamma(\alpha))} \Pi_{\sigma ,\alpha} \le \Pi_{\sigma ,\alpha}
{\sigma} \Pi_{\sigma ,\alpha}
\le 2^{-n(S(\sigma)-\gamma(\alpha))} \Pi_{\sigma ,\alpha}
\text{ .}\end{equation}

For 
$a^n \in {\mathcal{T}}^n_{P, \alpha}$ 
there are positive constants $\beta(\alpha)'$, $\gamma(\alpha)'$, 
and $\delta(\alpha)'$, depending on $\alpha$ and tending to zero
when $\alpha\rightarrow 0$  such that

\begin{equation} \label{te4}
\mathrm{tr}\left(\mathtt{V}^{\otimes n}(x^n) \Pi_{\mathtt{V}, \alpha}(x^n)\right)
> 1-2^{-n\beta(\alpha)'}
\text{ ,}\end{equation}

\begin{align} \label{te5}
&2^{-n(S(\mathtt{V}|P)+\gamma(\alpha)')} \Pi_{\mathtt{V}, \alpha}(x^n)
 \le \Pi_{\mathtt{V}, \alpha}(x^n)\mathtt{V}^{\otimes n}(x^n) \Pi_{\mathtt{V}, \alpha}(x^n)\notag\\
 &\le 2^{-n(S(\mathtt{V}|P)-\gamma(\alpha)')} \Pi_{\mathtt{V}, \alpha}(x^n)
\text{ ,}\end{align}
\begin{equation} \label{te6}
2^{n(S(\mathtt{V}|P)-\delta(\alpha)')}\le \mathrm{tr}\left(
\Pi_{\mathtt{V}, \alpha}(x^n) \right)\le 2^{n(S(\mathtt{V}|P)+\delta(\alpha)')}
\text{ .}\end{equation}

For the classical-quantum channel
$\mathtt{V}: \mathcal{X} \rightarrow \mathcal{S}({\cal H})$ and a probability
distribution $P$ on $\mathcal{X}$  we define
 a quantum state $P\mathtt{V}$ $:=$ $\sum_x P(x)\mathtt{V}(x)$ on $\mathcal{S}({\cal H})$.
 For $\alpha > 0$ we define an
orthogonal subspace projector $\Pi_{P\mathtt{V}, \alpha}$
 fulfilling (\ref{te1}), (\ref{te2}), and (\ref{te3}).
Let $x^n\in{\mathcal{T}}^n_{P, \alpha}$.
For $\Pi_{P\mathtt{V}, \alpha}$ there is a positive constant $\beta(\alpha)''$ such that
following inequality holds:
\begin{equation} \label{te7}  \mathrm{tr} \left(  \rho^{\otimes n}(x^n) \cdot \Pi_{P\mathtt{V}, \alpha } \right)
 \geq 1-2^{-n\beta(\alpha)''} \text{ .}\end{equation}\vspace{0.2cm}

We give here a sketch of the proof. For a detailed proof please see \cite{Wil}.\vspace{0.2cm}

\it{proof}

(\ref{te1}) holds because 
$\mathrm{tr}\left({\sigma} \Pi_{\sigma ,\alpha}\right)$
$=$ $\mathrm{tr}\left(\Pi_{\sigma ,\alpha}{\sigma} \Pi_{\sigma ,\alpha}\right) $
$=$   $P^n({\mathcal{T}}^n_{P, \alpha})$.  
(\ref{te2}) holds because
$\mathrm{tr} \left(\Pi_{\sigma ,\alpha}\right)$ $=$
$\left\vert {\mathcal{T}}^n_{P, \alpha} \right\vert$.
(\ref{te3}) holds because
$2^{-n(S(\sigma)+\gamma(\alpha))}$ $\le$
$P^n(x^n)$ $\le$ $2^{-n(S(\sigma)-\gamma(\alpha))}$
for $x\in {\mathcal{T}}^n_{P, \alpha}$ and a positive $\gamma(\alpha)$.
(\ref{te4}), (\ref{te5}), and (\ref{te6})
can be obtained in a similar way.
(\ref{te7}) follows from the permutation-invariance of $\Pi_{P\mathtt{V}, \alpha}$.
\begin{flushright}$\square$\end{flushright}\vspace{0.2cm}

\begin{definition}

A arbitrarily varying classical-quantum channel (AVCQC) ${\cal W}$ is specified by a 
set $\{\{\rho(x,s), x \in {\cal X}\}, s \in {\cal S}\}$ of classical quantum channels with a common input alphabet ${\cal X}$ and output space ${\cal H}$, 
which are indexed by elements $s$ in a finite set ${\cal S}$. Elements $s \in {\cal S}$ usually are called the states of the channel.
${\cal W}$ outputs a quantum state
\begin{equation} \label{eq_f1a}
\rho^{\otimes n}({\bf x}, {\bf s}):=\rho(x_1, s_1) \otimes \rho(x_2, s_2) \otimes \ldots, \otimes \rho(x_n, s_n),
\end{equation}
if an input codeword ${\bf x}=(x_1,x_2, \ldots, x_n)$ is input into the channel, 
and the channel is governed by a state sequence ${\bf s}=(s_1, s_2, \ldots, s_n)$,
 while the
state varies from symbol to symbol in an  arbitrary
manner.

\end{definition}

We assume that
the channel state $s$ is in control  of
the jammer.
Without loss of generality we also assume
that the jammer always chooses the most advantageous attacking strategy
according to 
his knowledge.

\begin{definition}
A code $\gamma :=({\cal U}, \{{\cal D}(i), i \in {\cal I}\})$ of length $n$ for a classical quantum channel consists of its code book
${\cal U}$ and decoding measurement $\{{\cal D}(i), i \in {\cal I}\}$, where the code book ${\cal U}:=\{{\bf u}(i), i \in {\cal I}\}$ is a subset of input alphabet ${\cal X}^n$ indexed by messages $i$ in the message set ${\cal I}$, and the decoding measurement $\{{\cal D}(i), i \in {\cal I}\}$ is a quantum measurement in the output space ${\cal H}^{\otimes n}$ that is, ${\cal D}(i) \ge 0$ for all $i \in {\cal I}$ and $\sum_{i \in {\cal I}}, {\cal D}(i)=\mathbb{I}_{\cal H}$. 
\end{definition}

\begin{definition}

A random correlated code $\Gamma$ for a AVCQC ${\cal W}$ is a uniformly distributed
random variable taking values in a set of codes 
$\{({\cal U}(k), \{{\cal D}(j,k), j \in {\cal J}\}), k \in {\cal K}\}$ with a common 
message set ${\cal J}$, where ${\cal U}(k)=\{{\bf u}(j,k), j \in {\cal J}\}$ and $\{{\cal D}(j,k), j \in {\cal J}\}$ 
are the code book and decoding measurement of the $k$th code in the set respectively. 
$|{\cal K}|$ is called the key size.
\end{definition}

\begin{remark} Usually
 a random correlated code is defined as
any random variable taking values in a set of codes.
Here we restrict ourselves to uniformly distributed
 random variables, since it is
sufficiently for our purpose (cf. \cite{Wi/No/Bo}).
\end{remark}

\textbf{\textit{C: Capacity definitions and basic relations}}

One of the fundamental task  of quantum Shannon theory is
to characterize  performance
 measurements
maximizing  the efficiency of 
quantum communication. Hence we 
introduce here capacity for message transmission
and simple relations between different
quantities.

\begin{figure}[H]\begin{center} \includegraphics[width=0.75\textwidth]{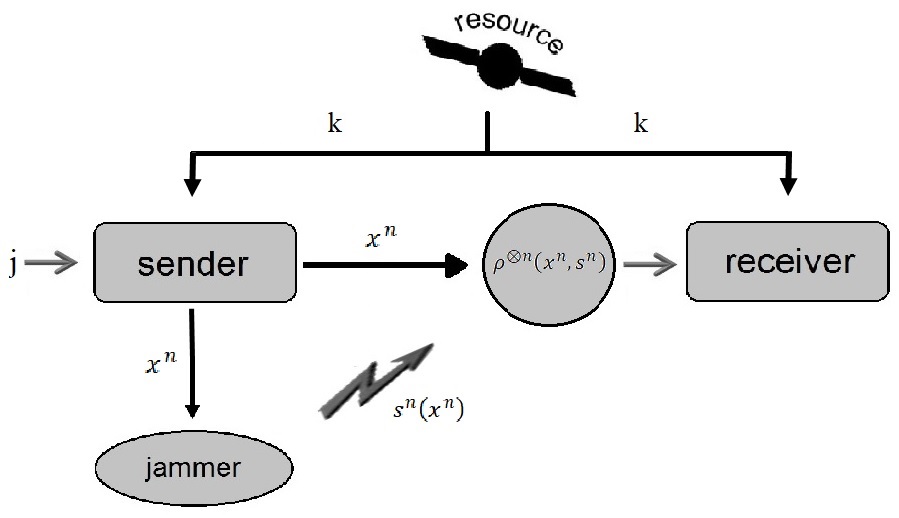}
  \caption{The jammer  knows both the coding scheme and the input
codeword (scenario 1)}\label{scenario1}
\end{center}\end{figure}

\begin{figure}[H]\begin{center}  
 \includegraphics[width=0.75\textwidth]{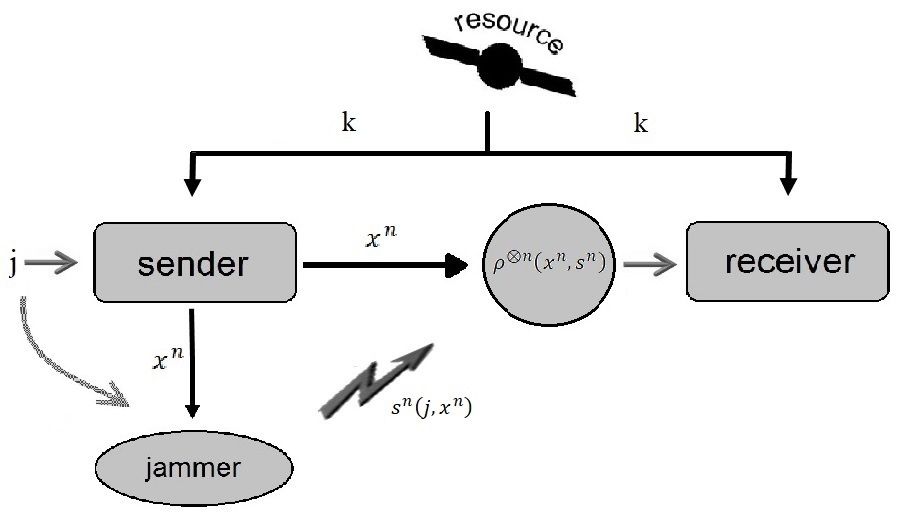}
  \caption{The jammer  knows    coding scheme,   input
codeword, and  message (scenario 2)}\label{scenario2}
\end{center}\end{figure}

As already mentioned this work concentrates on
message transmission over classical quantum
channels with a jammer with additonal  side information.
It is clear that this  side information are
encoded by the same  coding scheme, which 
is known by the jammer by assumption, as
the legal transmitters use for their communication.
We assume
that the jammer  chooses the most advantageous attacking strategy
according to 
his side information.
We now distinguish two scenarios depending on
the jammer's knowledge (cf. Figure \ref{scenario1}
and \ref{scenario2}). We consider for each scenario
both average and maximum error criteria.

\textbf{\textit{Scenario 1}}\rm

In this scenario
jammer knows coding scheme and input codeword but not the message to be sent.

\begin{definition}
By assuming that the random message $J$ is uniformly distributed, we define the average probability of error by
\begin{align} 
&p_a(\Gamma)\nonumber \\
&=\max_{{\bf s}}\mathbb{E}tr[\rho^{\otimes n}({\bf u}(J,K), {\bf s} ({\bf u}(J,K)))(\mathbb{I}_{\cal H}-{\cal D}(J,K))] \nonumber \\
&=\max_{{\bf s}}\frac{1}{|{\cal J}|} \sum_{j \in {\cal J}} \sum_{k \in {\cal K}} Pr\{K=k\}\nonumber \\
&tr[\rho^{\otimes n}({\bf u}(i,k), {\bf s} ({\bf u}(j,k)))(\mathbb{I}_{\cal H}-{\cal D}(j,k))].
\end{align}
This can be also rewritten as
\begin{align}\label{eq_f2a} &
p_a(\Gamma)\nonumber \\
&=\sum_{{\bf x}} Pr\{{\bf u}(J,K)={\bf x}\} \max_{{\bf s} \in {\cal S}^n}\mathbb{E}\{tr[\rho^{\otimes n}({\bf u}(J,K), {\bf s})\nonumber \\
&(\mathbb{I}_{\cal H}-{\cal D}(J,K))]|{\bf u}(J,K)={\bf x}\}.
\end{align}

The maximum probability of error  is defined as
\begin{align}\label{eq_f3} & 
p_m(\Gamma)\nonumber \\
&=\max_{j \in {\cal J}} \max_{{\bf s}}\mathbb{E}tr[\rho^{\otimes n}({\bf u}(j,K), {\bf s} ({\bf u}(j,K)))(\mathbb{I}_{\cal H}-{\cal D}(j,K))].
\end{align}

\end{definition}

\begin{definition}A non-negative number $R$ is an achievable \bf rate \it  for the
arbitrarily varying classical-quantum  channel
${\cal W}$ \bf under random correlated  coding in scenario 1
  under the average error criterion \it and \bf under the maximal error
criterion \it
  if  for every
 $\delta>0$ and  $\epsilon>0$, if $n$ is sufficiently large,
there is an 
 random correlated code
 $\Gamma$  of length
$n$ such that
$\frac{\log |{\cal J}|}{n} > R-\delta$, and
$p_a(\Gamma) < \epsilon$ and
$p_m(\Gamma) < \epsilon$, respectively.

 The
supremum on achievable rate under random correlated  coding   of ${\cal W}$ under the average 
error criterion and under the maximal error
criterion in scenario 1 is called the
random correlated    capacity of
 ${\cal W}$ 
under the average error criterion and under the maximal error
criterion in scenario 1, denoted by
$C^{*}({\cal W})$  and $C^{*}_m({\cal W})$, respectively.

\end{definition}

\begin{definition}
Let $\epsilon\in [0,1)$.
A non-negative number $R$ is an $\epsilon$ - achievable \bf rate \it  for the
arbitrarily varying classical-quantum  channel
${\cal W}$ \bf under random correlated  coding in scenario 1
  under the average error criterion \it and \bf under the maximal error
criterion \it
  if  for every
 $\delta>0$  if $n$ is sufficiently large,
there is an 
 random correlated code
 $\Gamma$  of length
$n$ such that
$\frac{\log |{\cal J}|}{n} > R-\delta$, and
$p_a(\Gamma) < \epsilon$ and
$p_m(\Gamma) < \epsilon$, respectively.

 The
supremum on achievable rate under random correlated  coding   of ${\cal W}$ under the average 
error criterion and under the maximal error
criterion in scenario 1 is called the
random correlated    $\epsilon$ - capacity of
 ${\cal W}$ 
under the average error criterion and under the maximal error
criterion in scenario 1, denoted by
$C^{*}({\cal W},\epsilon)$  and $C^{*}_m({\cal W},\epsilon)$, respectively.

\end{definition}

By (\ref{eq_f2a}) it is clear, that to employ a ``mixed strategy" for the jammer may not do better than only to use deterministic strategy. That is, 
the jammer may not enlarge the average probability of error, if he randomly chooses a state sequence with any
 conditional distribution $Q: {\cal X}^n \rightarrow {\cal S}^n$, according to the input codeword, 
instead chooses a fixed state sequence with the best deterministic strategy, because  
\begin{eqnarray}
&&\sum_{{\bf s} \in {\cal S}^n}Q({\bf s}|{\bf x})\mathbb{E}\{tr[\rho^{\otimes n}({\bf u}(J,K), {\bf s})(\mathbb{I}_{\cal H}-{\cal D}(J,K))]|{\bf u}(J,K) ={\bf x}\} \nonumber \\
&& \le \max_{{\bf s} \in {\cal S}^n}\mathbb{E}\{tr[\rho^{\otimes n}({\bf u}(J,K), {\bf s})(\mathbb{I}_{\cal H}-{\cal D}(J,K))]|{\bf u}(J,K)={\bf x}\} \nonumber
\end{eqnarray}
for all $Q$ and all ${\bf x}$ (with $ Pr\{{\bf u}(J,K)={\bf x}\}>0$).

\textbf{\textit{Scenario 2}}\rm

Now the jammer has more benefit and he can choose the state sequence according to both input codeword and message which sender wants to transmit, or a function $\psi$ $:\cup_{k \in {\cal K}} {\cal U}(k) \times {\cal J}$ 
 $\rightarrow {\cal S}^n$.

\begin{definition}
We define the average probability of error in scenario 2 by
\begin{align}  \label{eq_f4}
&p_a^{**}(\Gamma)=\max_{\psi}\sum_{j \in {\cal J}} \frac{1}{|{\cal J}|} \mathbb{E}tr[\rho^{\otimes n}\nonumber \allowdisplaybreaks\\
&({\bf u}(j,K), \psi ({\bf u}(j,K),j))(\mathbb{I}_{\cal H}-{\cal D}(j,K))].
\end{align}
The maximum probability of error in scenario 2  is defined as
\begin{align}\label{eq_f5} & 
p_m^{**}(\Gamma)=\max_{j \in {\cal J}}\max_{\psi}\mathbb{E}
tr[\rho^{\otimes n}\nonumber \allowdisplaybreaks\\
&({\bf u}(j,K), \psi ({\bf u}(j,K),j))(\mathbb{I}_{\cal H}-{\cal D}(j,K))].
\end{align}
\end{definition}

\begin{definition}A non-negative number $R$ is an achievable \bf rate \it  for the
arbitrarily varying classical-quantum  channel
${\cal W}$ \bf  under random correlated  coding in scenario 2 \bf under the average error criterion \it and \bf under the maximal error
criterion \it
  if  for every
 $\delta>0$ and  $\epsilon>0$, if $n$ is sufficiently large,
there is an 
 random correlated code
 $\Gamma$  of length
$n$   such that
 $\frac{\log |{\cal J}|}{n} > R-\delta$, and
 $p_a^{**}(\Gamma) < \epsilon$ and
$p_m^{**}(\Gamma) < \epsilon$, respectively.


 The
supremum on achievable rate under random correlated  coding   of ${\cal W}$ under the average 
error criterion and under the maximal error
criterion in scenario 2 is called the
random correlated   capacity of
 ${\cal W}$   
under the average error criterion and under the maximal error
criterion in scenario 2, denoted by
$C^{**}({\cal W})$  and $C^{**}_m({\cal W})$, respectively.
\end{definition}

\begin{definition}
Let $\epsilon\in [0,1)$.
A non-negative number $R$ is an $\epsilon$ - achievable \bf rate \it  for the
arbitrarily varying classical-quantum  channel
${\cal W}$ \bf  under random correlated  coding in scenario 2 \bf under the average error criterion \it and \bf under the maximal error
criterion \it
  if  for every
 $\delta>0$, if $n$ is sufficiently large,
there is an 
 random correlated code
 $\Gamma$  of length
$n$   such that
 $\frac{\log |{\cal J}|}{n} > R-\delta$, and
 $p_a^{**}(\Gamma) < \epsilon$ and
$p_m^{**}(\Gamma) < \epsilon$, respectively.


 The
supremum on  $\epsilon$ - achievable rate under random correlated  coding   of ${\cal W}$ under the average 
error criterion and under the maximal error
criterion in scenario 2 is called the
random correlated    $\epsilon$ - capacity of
 ${\cal W}$   
under the average error criterion and under the maximal error
criterion in scenario 2, denoted by
$C^{**}({\cal W},\epsilon)$  and $C^{**}_m({\cal W},\epsilon)$, respectively.
\end{definition}

Obviously
\[C^{**}({\cal W}) \le C^{*}({\cal W}).\]
It is easy to show that 
\[C^{*}_m({\cal W})= C^{**}_m({\cal W}),\]
because both (\ref{eq_f3}) and (\ref{eq_f5}) are equal to
\[\max_j\sum_{{\bf x}} Pr\{{\bf u}(j,K)={\bf x}\} \max_{{\bf s} \in {\cal S}^n}
\mathbb{E}\{tr[\rho^{\otimes n}({\bf u}(j,K), {\bf s})(\mathbb{I}_{\cal H}-{\cal D}(j,K))]| {\bf u}(j,K)={\bf x}\}.\]
Moreover,
the average probability of error (\ref{eq_f4}) can rewritten as
\[\sum_{j \in {\cal J}} \frac{1}{|{\cal J}|} \sum_{\bf x} Pr\{{\bf u}(j,K)={\bf x}\} 
\max_{{\bf s}\in {\cal S}^n}\mathbb{E}tr[\rho^{\otimes n}({\bf u}(j,K), {\bf s})(\mathbb{I}_{\cal H}-{\cal D}(J,K))|{\bf u}(j,K)={\bf x}].\]
Thus, in the standard way, by Markov inequality one may conclude that 
the message set ${\cal J}$ of any code with average probability of error $\lambda$ in scenario 2
contains a subset ${\cal J}'$ 
such that $|{\cal J}'| \ge \frac{|{\cal J}|}{2}$ and
\[\max_j\sum_{{\bf x}} Pr\{{\bf u}(j,K)={\bf x}\} \max_{{\bf s} \in {\cal S}^n}\mathbb{E}
\{tr[\rho^{\otimes n}({\bf u}(j,K), {\bf s})(\mathbb{I}_{\cal H}-{\cal D}(j,K))]| {\bf u}(j,K)
={\bf x}\} \le 2 \lambda,\] for all $j \in {\cal J}'$. 
That is, \[C^{**}({\cal W})= C^{**}_m({\cal W}),\]
thus
\begin{equation} \label{eq_cap}C^{*}({\cal W})\geq C^{*}_m({\cal W})=C^{**}({\cal W})= C^{**}_m({\cal W}).\end{equation}

\section{Main Results}
\label{MR}

For a given AVCQC ${\cal W}=\{\{\rho(x,s), x \in {\cal X}\}, s \in {\cal S}\}$ with set of state ${\cal S}$,  let
\begin{equation} \label{eq_doublebar}
\bar{\bar{{\cal W}}}:=\{\{\bar{\bar{\rho}}_{Q}( x):=\sum_s Q(s|x) \rho(x,s), x \in {\cal X}\}: \mbox{ for all $Q: {\cal X} \rightarrow {\cal S}$}\}.
\end{equation}

\begin{theorem} \label{thm_d} (Direct Coding Theorem  for Scenario 1)
  Given a AVCQC ${\cal W}=\{\{\rho(x,s), x \in {\cal X}\}, s \in {\cal S}\}$ and a type $P_X$ , for all $\epsilon >0$, and $\lambda >0$, there is a $b>0$, such that for all sufficiently large $n$, there exists a 
code $\Gamma$ of length $n$
  with a rate larger than $\min_{\bar{\bar{\rho}}(\cdot) \in \bar{\bar{W}}} 
\chi(P_X, \bar{\bar{\rho}}(\cdot))-\epsilon$, 
	average probability of error in scenario 1 smaller than $\lambda$, and key size of the random correlated code smaller then $bn^2$. Moreover 
	codewords of code books in support set of the random correlated code $\Gamma$, all are in ${\cal T}^n_X$.
  \end{theorem}
\begin{remark}
 In particular, there is a constant $a >0$ (depending only on the AVCQC) 
such that for any sequence of positive real numbers $\{\lambda_n\}$, lower bounded by
 $\lambda_n \geq 2^{-n \alpha}$ for an $\alpha >0$ (depending on $\epsilon$), 
with $\lim_{n \rightarrow \infty}\lambda_n=0$, there exists a  sequence of random 
correlated codes with a rate larger than $\min_{\bar{\bar{\rho}}(\cdot) \in \bar{\bar{W}}} \chi(P_X, \bar{\bar{\rho}}(\cdot))-\epsilon$,
 average probability of error smaller than $\lambda_n$ 
and        the amount of common randomness        upper bounded by $\frac{an^2}{\lambda_n^3}$.
  \label{iptiac}\end{remark}

  \begin{theorem}\label{thmc} (Strong Converse Coding Theorem  for Scenario 1)

	For every $\epsilon\in [0,1)$ we have
\begin{equation} \label{eq_r2}
C^*({\cal W},\epsilon) \le \max_P \min_{\bar{\bar{\rho}} (\cdot) \in \bar{\bar{\cal W}}} \chi(P, \bar{\bar{\rho}} (\cdot)).
\end{equation}
\end{theorem}\vspace{0.2cm}

Let
\begin{equation} \label{eq_bar}
\bar{{\cal W}}:=\{\{\bar{\rho}_P( x):=\sum_s P(s) \rho(x,s), x \in {\cal X}\}: \mbox{ for all probability distributions $P$ on ${\cal S}$}\}.
\end{equation}
Then obviously
\begin{equation} \label{eq_r2b}
\max_P \min_{\bar{\bar{\rho}} (\cdot) \in \bar{\bar{\cal W}}} 
\chi(P, \bar{\bar{\rho}} (\cdot)) \le \max_P \min_{\bar{\rho} (\cdot) 
\in \bar{\cal W}} \chi(P, \bar{\rho} (\cdot)).\end{equation}

The following Example  \ref{elcxcya013} shows that the inequality is strict 
already in classical  arbitrarily varying channels, as a special case of AVCQC. It was shown the 
 random 
correlated capacities  of a AVCQC under
maximum error probability and average error probability when the jammer does not know the 
channel input are the same and both equal to $\max_P 
\min_{\bar{\rho} (\cdot) \in \bar{\cal W}} \chi(P, \bar{\rho} (\cdot))$. 
Recalling that to employ the criterion of average probability of error 
corresponds to scenario 1
and 
the criterion of maximum probability of error corresponds to scenario 2, we conclude 
that knowing the message to be sent may not help a jammer who only know the coding scheme, 
for reduction the capacity, if random correlated codes 
are allowed to be used by the communicators side.

\begin{example}\label{elcxcya013}
Let ${\cal X} = {\cal Y} = \{ 0, 1\}$ and ${\cal S} = \{s_0, s_1\}$.
We define a  classical  arbitrarily varying channel $\cal W $
represented by the transmission
matrices 
\[\left( \begin{array} {rr}
\frac{3}{4}&\frac{1}{4}\\
\frac{1}{2}&\frac{1}{2}\\
 \end{array}\right)\text{ ,}
~~\left( \begin{array} {rr}
\frac{1}{4}&\frac{3}{4}\\
0&1\\
\end{array}\right)\text{ .}\]

The jammer may choose  $Q$ by setting
$Q(s_0|0) = Q(s_1|0) = \frac{1}{2}$, $Q(s_0|1) =1$
and $Q(s_1|1) =0$.
Since 
\[ \frac{1}{2} \cdot \left(\begin{array} {rr} 
\frac{3}{4}&\frac{1}{4}\end{array}
\right)
+ \frac{1}{2} \cdot \left( \begin{array}  {rr} 
\frac{1}{4}&\frac{3}{4}\end{array}
\right)
= 1\cdot \left ( \begin{array} {rr} 
\frac{1}{2}&\frac{1}{2}\end{array}
\right) +
0 \cdot \left(  
0, 1
 \right) ,\]
we have
\[C^*({\cal W})=0.\]

But when the jammer has no knowledge about
the channel  input, we can always achieve
positive capacity, since zero capacity means
there is a $a\in(0,1)$ such that
\[a\cdot \left ( \begin{array} {rr}
\frac{3}{4}&\frac{1}{4}\\
\frac{1}{2}&\frac{1}{2}\\
 \end{array}\right ) ~ +
~(1-a)\cdot\left ( \begin{array} {rr}
\frac{1}{4}&\frac{3}{4}\\
0&1\\
\end{array}\right )\]
has rank $1$, which can only be true
when 
\[ a \cdot \left(\begin{array} {rr} 
\frac{3}{4}&\frac{1}{4}\end{array}
\right)
+ (1-a) \cdot \left ( \begin{array} {rr} 
\frac{1}{4}&\frac{3}{4}\end{array}
\right)
=a \cdot \left( \begin{array}  {rr} 
\frac{1}{2}&\frac{1}{2}\end{array}
\right)  +
(1-a) \cdot \left(  
0, 1
 \right) .\]

But there is clearly no such 
$a\in(0,1)$ since else we would have
\begin{align*}&\frac{3}{4}a+\frac{1}{4}(1-a) = \frac{1}{2}a\\
& \Rightarrow \frac{1}{4} =\frac{1}{2}a+\frac{1}{4}a-\frac{3}{4}a\\
& \Rightarrow \lightning .\end{align*}

Thus  when the jammer has no knowledge about
the channel  input, this channel has a positive deterministic capacity.
\end{example}

Example \ref{elcxcya013} shows that the jammer really benefits
from his knowledge about the channel input.\vspace{0.2cm}

The following example was first presented at the  IEEE International Symposium on Information Theory 2010
in a talk by N. Cai, T. Chen, and A, Grant.

\begin{example}\label{elcxcya012}
Let ${\cal X} = {\cal Y} = \{a, 0, 1, 2\}$ and ${\cal S} = \{s_0, s_1\}$.
We define a  classical  arbitrarily varying channel $\cal W $
such that $W(a|a, s_0) = W(a|a, s_1) = 1$, $W(y|x; s_i) = 1$ if
$y = x+i~(mod3)$  for $x, y \in \{0, 1, 2\}$. That is the transmission
matrices in $\cal W $ are
\[\left ( \begin{array} {rrrr}
1&0&0&0\\
0&1&0&0\\
0&0&1&0\\
0&0&0&1\\ \end{array}\right )\text{ ,}
~~\left ( \begin{array} {rrrr}
1&0&0&0\\
0&0&1&0\\
0&0&0&1\\
0&1&0&0\\ \end{array}\right )\text{ .}\]

At first we have that the deterministic  capacity  of  $W$ under
maximum error probability is larger or equal to  $2$ because for all $n, \{a, 0\}^n$
there
is a zero-error code of length $n$ and therefore a code with
criterion of maximum probability of error. Secondly let $g$ be
a mapping from ${\cal X}^n \rightarrow \{a, 0\}^n$  for arbitrary $n$ sending $x^n$
to $y^n$ such that $y_i = a$ if $x_i = a$ and otherwise $x_i = 0$, for
$i = 1, 2, \cdots , n$. Then no pair of codewords in   a code with
criterion of maximum probability of error have the same image
under the mapping $g$ because in probability one the decoder
may not separate the two codewords with the same image if
the jammer properly chooses the state sequence according to
the input codeword. Thus the deterministic  capacity  of $W$ under
maximum error probability is  equal to  $2$.

On other hand let $P$ be a input
distribution such that $P(a) = \frac{2}{5}$
 and $P(i) = \frac{1}{5}$
for $i = 0, 1, 2$.
Let $X$ and $Y$ be the input and output random variables
for $P$ and $\bar{\bar{W}}$, the channel in $\bar{\bar{{\cal W}}}$,
 minimizing $I(P; \bar{\bar{W}})$.
 Then
$H(X) = \frac{2}{5}\log \frac{5}{2}
 +\frac{3}{5}\log 5$. Next by considering the support
sets of conditional distributions, we have $H(X|Y = a) = 0$
and $H(X|Y = i)\leq 1$ for $i = 0, 1, 2$. Thus $H(X|Y ) \leq \frac{3}{5}$
and
therefore $I(X; Y ) = H(X) - H(X|Y ) = \log \frac{5}{2}$. Moreover
by simple calculation, $I(P;W) = \log \frac{5}{2}$ for
$W(\cdot|\cdot) := \frac{1}{2}W(\cdot|\cdot,s_0)+\frac{1}{2}W(\cdot|\cdot,s_1)$. 
Thus $\min_{\bar{\bar{W}}\in{\bar{\bar{{\cal W}}}}} I(P; \bar{\bar{W}} )= \log \frac{5}{2}$.
and $ \max_{P\in P({\cal X})}
\min_{\bar{\bar{W}}\in{\bar{\bar{{\cal W}}}}} I(P; \bar{\bar{W}} )\geq \log \frac{5}{2}$.
\end{example}
Example \ref{elcxcya012} show that the legal transmitters really benefits
from  the resource even when the deterministic capacity under the maximal error criterion
is positive. \vspace{0.2cm}

Now one may concern the same question in scenario 2. 
This is answered by the following Theorem, which can be proven by modifying the proof of Theorem \ref{thm_d}:
\begin{theorem} \label{thm_da}
The same conclusion  for scenario 2, as that for scenario 1 in Theorem\ref{thm_d}, holds.
\end{theorem}

The above three Theorems 
and the facts that 
\[C^{*}({\cal W})\leq C^{*}({\cal W},\epsilon),
~~~ C^{*}_m({\cal W}) \leq C^{*}_m({\cal W},\epsilon) \leq C^{*}({\cal W},\epsilon),\]
\[C^{**}({\cal W}) \leq C^{**}({\cal W},\epsilon) \leq C^{*}({\cal W},\epsilon),
~~~ C^{**}_m({\cal W}) \leq C^{**}_m({\cal W},\epsilon) \leq C^{*}({\cal W},\epsilon),\]
 yield the coding theorem:
\begin{corollary} For all $\epsilon\in [0,1)$ we have
\begin{align}\label{eq_r1}
&C^{*}({\cal W})=C^{**}({\cal W}) = C^{*}_m({\cal W})= C^{**}_m({\cal W})\notag\\
&=C^{*}({\cal W},\epsilon)=C^{**}({\cal W},\epsilon) = C^{*}_m({\cal W},\epsilon)= C^{**}_m({\cal W},\epsilon)\notag\\
&= \max_P \min_{\bar{\bar{\rho}} (\cdot) \in \bar{\bar{\cal W}}} \chi(P, \bar{\bar{\rho}} (\cdot)).
\end{align}
Moreover the both capacity $C^{**}({\cal W})$ and $C^*({\cal W})$ can be achieved by codes with vanishing key rates.
\end{corollary}

Thus we conclude that:
\begin{itemize}
\item Further knowing message to be sent, may help a jammer to reduce the capacity neither in the scenario that the jammer knows coding scheme nor in the scenario that the jammer knows both coding scheme and input codeword.
\item knowing input codeword is more effectual than knowing the message for a jammer, who knows coding scheme, for attack the communication.
\end{itemize}

\section{proof Theorem \ref{thm_d}} \label{sec_prd}

Although coding for classical arbitrarily varying channels is already a 
challenging topic with a lot of open problems, coding for AVCQC is even much  
harder. Due to the non-commutativity of quantum operators, 
many
techniques, concepts and methods of classical information theory,
for instance, non-standard decoder and list decoding,
may not be extended to quantum information theory. 
Sarwate  used in \cite{Sar} list decoding 
to prove the coding theorem for   classical
arbitrarily varying 
 channels when the jammer knows input codeword. 
However since how to apply list decoding for
quantum channels is still an open problem,
the technique for classical  channels in \cite{Sar}
 can not be extended to  AVCQC.  
We need a different approach for our scenario 1.

If the jammer would have some information about the outcome $k$ of the random key 
through the input codeword, to which he has access in  scenario 1,
he could apply a strategy against the $k$th 
deterministic coding for AVCQC
by choosing the worst state sequence to attack the communication, which we do not want.
To this end a codeword  must be used by ``many" outcomes $\gamma (k)$ of 
a random correlated code $\Gamma$, if it is used by at least one of $\gamma (k)$. 
This is the main idea of our proof.
We divide the proof into 5 steps. At the first step we derive a useful auxiliary result from known results. Next with the auxiliary result and Chernoff bound, we shall generate a ground set of code books from a typical set ${\cal T}^n_X$. Then our code $\Gamma$ is constructed through the ground set and analyzed at the 3th and 4th steps, respectively. To simplify the statement, we shall not fix the values of parameters at the 2-4th steps exactly, but only set up necessary constraints to them. So finally we have to assign values to the parameters appearing in the proof at the last step.

\subsection{An Auxiliary Result}

We first derive a useful auxiliary result from known projections in previous work.

To construct decoding measurements of codes for classical quantum compound channel
the authors in \cite{BB} and \cite{H} introduced two kinds of projections for a set of classical quantum channels and input codewords ${\bf x} \in {\cal T}^n_X$ respectively. Although the two projections are quite different, they share the same properties. We summary their properties, which will be used in the paper, as the following lemma.
\begin{lemma} \label{lemma_comp}
For a set of classical quantum channels $\tilde{\cal W}$ with a common input alphabet ${\cal X}$ and a common output Hilbert space ${\cal H}$
and any an input codeword ${\bf x} \in {\cal T}^n_X$, there exits a projection ${\cal P}({\bf x})$ in ${\cal H}$ such that,

(i) For all
$\tilde{\rho}(\cdot) \in \tilde{\cal W}$,
\begin{equation} \label{eq_d1}
tr(\tilde{\rho}^{\otimes n}({\bf x}){\cal P}({\bf x})) > 1-2^{-n\eta}
\end{equation}
for an $\eta >0$;

(ii)
\begin{equation} \label{eq_d2}
tr (\tilde{\rho}_X^{\otimes n} {\cal P}({\bf x})) < 2^{-n[\min_{\tilde{\rho}(\cdot) \in \tilde{\cal W}} \chi(P_X, \tilde{\rho}(\cdot))-\nu]},
\end{equation}
for all $\nu >0$, $\tilde{\rho}(\cdot) \in \tilde{\cal W}$
and sufficiently large $n$, where
\[\tilde{\rho}_X:=\sum_{x \in {\cal X}} P_X(x) \tilde{\rho}(x).\]
(iii) Moreover, for all permutation $\pi$ on $[n]=\{1,2,\ldots, n\}$ with ${\bf x}=(x_1,x_2, \ldots,x_n)=(x_{\pi(1)},x_{\pi(2)}, \ldots,x_{\pi(n)})$, ${\cal P}({\bf x})$ keeps invariant when permutation $\pi$ acts on coordinates of $n$th tensor power ${\cal H}^n$ of Hilbert space ${\cal H}$.
\end{lemma}

Let ${\cal W}=\{\rho (\cdot, s)=\{\rho(x, s), x \in {\cal X}\}, s \in {\cal S}\}$ be a finite set of classical quantum channels, indexed by elements of ${\cal S}$ and let $\bar{\bar{\cal W}}$ is defined by (\ref{eq_doublebar}). Then
\begin{corollary} \label{cor_0}
Let ${\cal P}({\bf x}')$ be the projection in Lemma \ref{lemma_comp} for $\tilde{\cal W}=\bar{\bar{\cal W}}$,
${\bf x} \in {\cal T}^n_{X}, {\bf s} \in {\cal S}^n$ and ${\bf X}'$
be randomly and uniformly distributed on ${\cal T}^n_X$, then
\begin{equation}\label{eq_d3}
\mathbb{E} tr (\rho^{\otimes n}({\bf x},{\bf s}) {\cal P}({\bf X}')) < 2^{-n[\min_{\bar{\bar{\rho}}(\cdot) \in \bar{\bar{\cal W}} }\chi(P_X, \bar{\bar{\rho}}(\cdot))-\nu-\xi]},
\end{equation}
for all $\xi>0$ and sufficiently large $n$.
\end{corollary}
{\it Proof:} Let $P_{XS}$ be joint type of $({\bf x},{\bf s})$. Let $({\bf X}, {\bf S})$ be randomly and uniformly distributed on ${\cal T}^n_{XS}$ and ${\bf X}'$ be random variable with uniform distribution on ${\cal T}^n_X$, and independent of  $({\bf X}, {\bf S})$. Then by Lemma \ref{lemma_comp} (ii), we have that
\begin{eqnarray}\label{eq_d4}
&&\mathbb{E} tr (\rho^{\otimes n} ({\bf X},{\bf S}){\cal P}({\bf X}'))\nonumber \\
&&=\sum_{{\bf x}' \in {\cal T}^n_X} Pr({\bf X}'={\bf x}')\sum_{({\bf x}, {\bf s}) \in {\cal T}^n_{XS}} Pr[({\bf X},{\bf S})=({\bf x},{\bf s})]tr[\rho^{\otimes n}({\bf x},{\bf s}){\cal P}({\bf x}')] \nonumber \\
&&<\sum_{{\bf x}' \in {\cal T}^n_X} Pr({\bf X}'={\bf x}')2^{n\xi}\sum_{{\bf x}\in {\cal X}^n {\bf s}\in {\cal S}^n} P_{XS}^n({\bf x},{\bf s})]tr[\rho^{\otimes n}({\bf x},{\bf s}){\cal P}({\bf x}')] \nonumber \\
&&=2^{n\xi}\sum_{{\bf x}' \in {\cal T}^n_X} Pr({\bf X}'={\bf x}')tr\{[\sum_{{\bf x}\in {\cal X}^n {\bf s}\in {\cal S}^n} \prod_{t=1}^nP_{XS}(x_t, s_t)\bigotimes_{t=1}^n\rho(x_t,s_t)]{\cal P}({\bf x}')\}\nonumber \\
&&=2^{n\xi}\sum_{{\bf x}' \in {\cal T}^n_X} Pr({\bf X}'={\bf x}') tr\{
[\sum_{x\in {\cal X}}P_X(x)
(\sum_{s \in {\cal S}} P_{S|X}(s|x)\rho(x,s))]^{\otimes n}{\cal P}({\bf x}')\}\nonumber \\
&&<2^{n\xi}
\sum_{{\bf x}' \in {\cal T}^n_X} Pr({\bf X}'={\bf x}')2^{-n[\min_{\bar{\bar{\rho}}(\cdot) \in \bar{\bar{\cal W}}} \chi(P_X, \bar{\bar{\rho}}(\cdot ))-\nu]}\nonumber \\
&&=2^{-n[\min_{\bar{\bar{\rho}}(\cdot) \in \bar{\bar{\cal W}}} \chi(P_X, \bar{\bar{\rho}}(\cdot))-\nu-\xi]},
\end{eqnarray}
for ${\bf x}=(x_1, x_2,\ldots,x_n)$ and ${\bf s}=(s_1,s_2, \ldots, s_n)$, where the first inequality holds because \[Pr[({\bf X},{\bf S})=({\bf x},{\bf s})]=\frac{1}{|{\cal T}^n_{XS}|} < 2^{-n(H(X,S)-\frac{\xi}{2})}<2^{n\xi}P_{XS}^n({\bf x},{\bf s})\]
 for all $\xi>0$ and sufficiently large $n$, if $({\bf x}, {\bf s}) \in {\cal T}^n_{XS}$, and equal to zero otherwise; and by (\ref{eq_d2}) the last inequality holds, because by (\ref{eq_doublebar}), $\{\sum_{s \in {\cal S}} P_{S|X}(s|x)\rho(x,s), x \in {\cal X}\} \in \bar{\bar{{\cal W}}}$.

Now by Lemma \ref{lemma_comp} (iii), we note that for all $({\bf x},{\bf s}) \in {\cal T}^n_{XS}, {\bf x}' \in {\cal T}^n_X$, the value of $tr[\rho^{\otimes n}({\bf x},{\bf s}) {\cal P}({\bf x}')]$ depends only on the joint type of $({\bf x},{\bf x}',{\bf s})$, and therefore for all $({\bf x},{\bf s}) \in {\cal T}^n_{XS}$, the value of
\[\sum_{{\bf x}'\in {\cal T}^n_X}Pr({\bf X}'={\bf x}')tr[\rho^{\otimes n}({\bf x},{\bf s}) {\cal P}({\bf x}')]\]
is a constant (only depending on the joint type of ({\bf x},{\bf s}). Thus (\ref{eq_d3}) follows from (\ref{eq_d4}) and the fact that
\[\mathbb{E} tr (\rho^{\otimes n} ({\bf X},{\bf S}){\cal P}({\bf X}'))=\sum_{({\bf x}, {\bf s}) \in {\cal T}^n_{XS}} Pr[({\bf X},{\bf S})=({\bf x},{\bf s})]\{\sum_{{\bf x}'\in {\cal T}^n_X}Pr({\bf X}'={\bf x}')tr[\rho^{\otimes n}({\bf x},{\bf s}) {\cal P}({\bf x}')]\}.\]
Thus, the proof is completed.

\subsection{Generation Ground Set for Code books} \label{subs_gr}

Let
\begin{equation} \label{eq_A}
A_n \ge 2^{-n[\min_{ \bar{\bar{\rho}}(\cdot) \in \bar{\bar{\cal W}}} \chi(P_X, \bar{\bar{\rho}}(\cdot))-\nu-\xi]}
\end{equation}
and ${\cal I}_n$ be a finite index set with the cardinality
\begin{equation} \label{eq_I}
|{\cal I}_n|> \frac{n\log_e|{\cal X}||{\cal S}|}{(3-e)A_n},
\end{equation}
which will be specified in Subsection \ref{subs_par}. Let ${\bf X}(i), i \in {\cal I}_n$ be randomly, independently and uniformly distributed on ${\cal T}^n_X$. Then by Corollary \ref{cor_0} and Chernoff bound, we have that for all ${\bf x} \in {\cal T}^n_{X}, {\bf s} \in {\cal S}^n$
\begin{eqnarray}\label{eq_d5}
&&Pr\{\sum_{i\in {\cal I}_n}tr[\rho^{\otimes n}({\bf x}, {\bf s}){\cal P}({\bf X}(i))] > 3A_n I_n\}  \nonumber \\
&&=Pr\{\exp_e [-3A_n I_n+\sum_{i \in{\cal I}_n}tr[\rho^{\otimes n}({\bf x}, {\bf s}){\cal P}({\bf X}(i))]] >1\}  \nonumber \\
&& \le e^{-3A_n I_n}\prod_{i \in {\cal I}_n} \mathbb{E}e^{tr[\rho^{\otimes n}({\bf x}, {\bf s}){\cal P}({\bf X}(i))]}
\nonumber \\
&& \le e^{-3A_n I_n}\prod_{i \in {\cal I}_n}[1+e\mathbb{E} \rho^{\otimes n}({\bf x}, {\bf s}){\cal P}({\bf X}(i))] \nonumber \\
&& \le  e^{-3A_n I_n}[1+e A_n]^{|{\cal I}_n|} \nonumber \\
&& \le \exp_e \{-3A_n |{\cal I}_n|+e A_n |{\cal I}_n|\}=e^{-(3-e) A_n |{\cal I}_n|},
\end{eqnarray}
where the first inequality is Chernoff bound, the second inequality holds because $e^z$ is a monotone increasing and convex function and so $e^z \le 1+ez$ for $z \in (0,1)$; the third inequality holds by (\ref{eq_d3}) and (\ref{eq_A}); and the last inequality follows from inequality $1+z \le e^z$. Thus by union bound and (\ref{eq_I}), we obtain that
\[Pr\{\cup_{{\bf x} \in {\cal T}^n_X,{\bf s} \in {\cal S}^n} [\sum_{i \in {\cal I}_n}tr[\rho^{\otimes n}({\bf x}, {\bf s}){\cal P}({\bf X}(i))] > 3A_n |{\cal I}_n|]\}< |{\cal X}|^n|{\cal S}|^n e^{-(3-e) A_n |{\cal I}_n|}<1.\]
Consequently we have that there exists a subset ${\cal B}=\{{\bf x}(i), i \in {\cal I}_n]\} \subset {\cal T}^n_X$, with
\begin{equation} \label{eq_d6}
\sum_{{\bf x}(i) \in {\cal B}}tr[\rho^{\otimes n}({\bf x}, {\bf s}){\cal P}({\bf x}(i))] \le 3A_n |{\cal I}_n|,
\end{equation}
for all ${\bf x} \in {\cal T}^n_X, {\bf s} \in {\cal S}^n$.

\subsection{Construction of Code} \label{subs_con}
\subsubsection{Generation of Code books}

Let ${\cal J}_n$ and ${\cal K}_n$ be two finite set and their cardinalities (depending on $n$) will be specified in Subsection \ref{subs_par}, but at this moment, we only assume that
\begin{equation} \label{eq_J1}
|{\cal J}_n| \le A_n^{-1}.
\end{equation}
Let
$({\bf U}(j,k),j \in {\cal J}_n), k \in {\cal K}_n$ be randomly uniformly and independently generated from
\[\{({\bf x}(i_1),{\bf x}(i_2),\ldots, {\bf x}(i_{|{\cal J}_n|})): i_j \in {\cal I}_n, \mbox{ for $j=1,2, \ldots, |{\cal J}_n|,$ with }
i_j \not= i_{j'}\mbox{ for } j \not= j'\}.\]
Then by (\ref{eq_d6}) we have that for all $i \in {\cal I}_n, {\bf s} \in {\cal S}^n,j, j' \in {\cal J}_n,$ with $ j \not= j'$ and $k \in {\cal K}_n$
\begin{eqnarray} \label{eq_d7}
&&\mathbb{E}tr[\rho^{\otimes n} ({\bf U}(j,k)), {\bf s}) {\cal P}({\bf U}(j',k))|{\bf U}(j,k))={\bf x}(i)]\nonumber \\
&&=\sum_{i' \in {\cal I}_n\setminus \{i\}} Pr[{\bf U}(j',k)={\bf x}(i')|{\bf U}(j,k))={\bf x}(i)]tr[\rho^{\otimes n} ({\bf x}(i), {\bf s}) {\cal P}({\bf x}(i'))]  \nonumber \\
&&=\frac{1}{|{\cal I}_n|-1}\sum_{i' \in {\cal I}_n\setminus \{i\}} tr[\rho^{\otimes n} ({\bf x}(i), {\bf s}) {\cal P}({\bf x}(i'))]\nonumber \\
&& \le \frac{1}{|{\cal I}_n|-1}\sum_{i' \in {\cal I}_n} tr[\rho^{\otimes n} ({\bf x}(i), {\bf s}) {\cal P}({\bf x}(i'))]\le \frac{3A_n|{\cal I}_n|}{|{\cal I}_n|-1}.
\end{eqnarray}
Consequently by Markov inequality we have that
\begin{eqnarray} \label{eq_mar}
&&Pr\{\sum_{j' \in {\cal J}_n\setminus \{j\}} tr[\rho^{\otimes n} ({\bf U}(j,k), {\bf s}) {\cal P}({\bf U}(j',k))] >\mu_n |{\bf U}(j,k)={\bf x}(i)\}]\nonumber \\
&& \le \frac{\mathbb{E}\{\sum_{j' \in {\cal J}_n\setminus \{j\}} tr[\rho^{\otimes n} ({\bf U}(j,k), {\bf s}) {\cal P}({\bf U}(j',k))] |{\bf U}(j,k)={\bf x}(i)\}}{\mu_n}\nonumber \\
&& = \frac{\sum_{j' \in {\cal J}_n\setminus \{j\}}\mathbb{E}\{ tr[\rho^{\otimes n} ({\bf U}(j,k), {\bf s}) {\cal P}({\bf U}(j',k))] |{\bf U}(j,k)={\bf x}(i)\}}{\mu_n}\nonumber \\
&&\le \frac{3A_n(|{\cal J}_n|-1)||{\cal I}_n|}{(|{\cal I}_n|-1)\mu_n}< \frac{3A_n|{\cal J}_n|}{\mu_n}
\end{eqnarray}
for all $i \in {\cal I}_n, {\bf s}\in {\cal S}^n, j \in {\cal J}_n, k \in {\cal K}_n$ and $\mu_n \in (0,1)$, where the last inequality holds because by (\ref{eq_I}) and (\ref{eq_J1}), $|{\cal J}_n| < |{\cal I}_n|$ and therefore $\frac{|{\cal J}_n|-1}{|{\cal I}_n|-1} < \frac{|{\cal J}_n|}{|{\cal I}_n|}$.
Therefore
\begin{eqnarray} \label{eq_d8}
&&Pr\{{\cal E}(i,{\bf s},k;\mu_n)\}= \sum_{j\in {\cal J}_n}Pr ({\bf U}(j,k)={\bf x}(i))Pr\{\sum_{j' \in {\cal J}_n\setminus \{j\}} tr[\rho^{\otimes n} ({\bf x}(i), {\bf s}) {\cal P}({\bf U}(j',k))] >\mu_n|{\bf U}(j,k)={\bf x}(i)\} \nonumber \\
&&< \frac{3A_n|{\cal J}_n|^2}{|{\cal I}_n|\mu_n},
\end{eqnarray}
for all $i \in {\cal I}_n, {\bf s} \in {\cal S}, k\in {\cal K}_n$ and $\mu_n \in (0,1)$, if we define ${\cal E}(i,{\bf s},k;\mu_n)$ as the random event that there exists a $j \in {\cal J}_n$ such that ${\bf U}(j,k)={\bf x}(i)$ and
\[\sum_{j' \in {\cal J}_n\setminus \{j\}} tr[\rho^{\otimes n} ({\bf x}(i), {\bf s}) {\cal P}({\bf U}(j',k))] >\mu_n.\]
In the sequel, we shall use the following version of well known Chernoff Bound.
\begin{lemma} \label{lemma_chernoff} ({\it Chernoff Bound}) Let
$B_1,B_2, \dots, B_L$ be i.i.d. random binary sequence taking values in $\{0,1\}$, with
$Pr(B_l=1)=p$. Then for all $\alpha \in (0,1), p_0\le p\le p_1$
\begin{equation}\label{eq_lemmach1}
Pr\{\sum_{l=1}^LB_l > Lp_1(1+\alpha)\}
<e^{-\frac{{\alpha}^2}{8}Lp_1},
\end{equation}
and
\begin{equation}\label{eq_lemmach2}
Pr\{\sum_{l=1}^LB_l < Lp_0(1-\alpha)\}
<e^{-\frac{{3\alpha}^2}{8}Lp_0}.
\end{equation}
\end{lemma}
For self-contained we prove it in Appendix \ref{sec_app}, although (\ref{eq_lemmach1}) was shown in \cite{C13} and (\ref{eq_lemmach2}) can be shown in a similar way.

Next for a fixed $i \in {\cal I}_n$, we define  random sets
\begin{equation} \label{eq_rset1}
\mathfrak{K}(i):=\{(k: \mbox{ there exists a $j \in {\cal J}_n$ with ${\bf U}(j,k)={\bf x}(i)$}\}
\end{equation}
and for all ${\bf s} \in {\cal S}^n$,
\begin{equation} \label{eq_rset12}
\mathfrak{K}_0(i,{\bf s})):=\{k: \mbox{ there exists a $j$ with ${\bf U}(j,k)={\bf x}(i)$ and  $\sum_{j' \in {\cal J}_n\setminus \{j\}} tr[\rho^{\otimes n} ({\bf x}(i), {\bf s}) {\cal P}({\bf U}(j',k))] >\mu_n$}\}.
\end{equation}
Let $\iota ({\cal E}(i,{\bf s}, k;\mu_n))$ be the indicator of the random event of ${\cal E}(i,{\bf s},k;\mu_n)$ (i.e., $\iota ({\cal E}(i,{\bf s},k;\mu_n))=1$ if ${\cal E}(i,{\bf s},k;\mu_n)$ occurs and otherwise $\iota ({\cal E}(i,{\bf s},k;\mu_n))=0$) and random variables
\[Z_i(k)=\left\{\begin{array}{ll} 1 & \mbox{if exists a $j$ with ${\bf U}(j,k) ={\bf x}(i)$} \\
                                     0 & \mbox{else.} \end{array}
                                     \right. \]
Then by (\ref{eq_d8}) we have that $Pr(\iota({\cal E}(i,k;\mu_n))=1)< \frac{3A_n|{\cal J}_n|^2}{|{\cal I}_n|\mu_n}$. By the definition of $Z_i(k)$
 we have that \[Pr(Z_i(k)=1)=\sum_{j\in {\cal J}_n} Pr[{\bf U}(j,k)={\bf x}(i)]= \frac{|{\cal J}_n|}{|{\cal I}_n|},\]
 as by the definition of ${\bf U}(j,k)$'s, the random events $\{{\bf U}(j,k)={\bf x}(i)\}, j \in {\cal J}_n$ are pairwise disjoint.

For each fixed $i \in {\cal I}_n$, we apply (\ref{eq_lemmach2}) to $[L]={\cal K}_n, B_k=Z_i(k), k \in {\cal K}_n$ and $p_0=\frac{|{\cal J}_n|}{|{\cal I}_n|}$ and obtain that
\begin{eqnarray} \label{eq_d9}
&&Pr\{|\mathfrak{K}(i)|<\frac{|{\cal K}_n||{\cal J}_n|}{|{\cal I}_n|}(1-\alpha) \} \nonumber \\
&&=Pr\{\sum_{k \in {\cal K}_n}Z_i(k) <|{\cal K}_n|\frac{|{\cal J}_n|}{|{\cal I}_n|}(1-\alpha)\}  \nonumber \\
&&< \exp_e\{-\frac{3 \alpha^2}{8}\frac{|{\cal K}_n||{\cal J}_n|}{|{\cal I}_n|}\}.
\end{eqnarray}
Similarly, by apply (\ref{eq_lemmach1}) to $[L]={\cal K}_n, B_k=\iota ({\cal E}(i,{\bf s},k;\mu)),k \in {\cal K}_n$ and $p_1=\frac{3A_n|{\cal J}_n|^2}{|{\cal I}_n|\mu_n}$, we have that
\begin{eqnarray} \label{eq_d10}
&&Pr\{|\mathfrak{K}_0(i,{\bf s}))|>\frac{3A_n|{\cal J}_n|^2 |{\cal K}_n|}{|{\cal I}_n|\mu_n}(1+\alpha)\} \nonumber \\
&&=Pr\{\sum_{k \in {\cal K}_n}\iota ({\cal E}(i,{\bf s},k;\mu_n))>|{\cal K}_n|\frac{3A_n|{\cal J}_n|^2}{|{\cal I}_n|\mu_n}(1+\alpha)\} \nonumber \\
&& <\exp_e\{-\frac{\alpha^2}{8}\frac{3A_n|{\cal J}_n|^2 |{\cal K}_n|}{|{\cal I}_n|\mu_n}\},
\end{eqnarray}
for all $i \in {\cal I}_n, {\bf s} \in {\cal S}$ and $\mu_n \in (0,1)$. Now choose $\alpha=\frac{1}{2}$, $|{\cal J}_n|$ and $\mu_n$ properly such that (\ref{eq_J1}) holds and
\begin{equation} \label{eq_AJ}
\lambda'_n:=\frac{A_n|{\cal J}_n|}{\mu_n} <1
\end{equation}
sufficiently small, $|{\cal K}_n|$ sufficiently large such that
\begin{equation} \label{eq_JK}
\frac{3}{32} \lambda'_n \frac{|{\cal K}_n||{\cal J}_n|}{|{\cal I}_n|} > 2n \log_e |{\cal S}||{\cal X}|,
\end{equation}
and
\begin{equation} \label{eq_Ia}
|{\cal I}_n| <|{\cal X}|^n,
\end{equation}
(all to be specified in Subsection \ref{subs_par})). Thus, by the union bound and (\ref{eq_d9}), (\ref{eq_d10}), (\ref{eq_AJ}), (\ref{eq_JK}) and (\ref{eq_Ia}), we have that
\[Pr\{ \cup_{i \in {\cal I}_n}[|\mathfrak{K}(i)|<\frac{|{\cal K}_n||{\cal J}_n|}{2|{\cal I}_n|}] \} < \frac{1}{2}, \]
and
\[Pr\{\cup_{{\bf s} \in {\cal S}}  \cup_{i \in {\cal I}_n}[|\mathfrak{K}_0(i,{\bf s}))|>\frac{9|{\cal J}_n||{\cal K}_n| \lambda'_n}{2|{\cal I}_n|}\} < \frac{1}{2} \] respectively. Consequently
\begin{equation} \label{eq_d11}
Pr\{[\cap_{i \in {\cal I}_n}(|\mathfrak{K}(i)|\ge \frac{|{\cal K}_n| |{\cal J}_n|}{2|{\cal I}_n|})]\cap [\cap_{{\bf s} \in {\cal S}}  \cap_{i \in {\cal I}_n}(|\mathfrak{K}_0(i,{\bf s})|\le \frac{9|{\cal J}_n||{\cal K}_n| \lambda'_n}{2|{\cal I}_n|})] \} >0
\end{equation}
Thus $\{{\bf U}(j,k) \in {\cal J}_n\}, k \in {\cal K}_n$
has a realization
\[{\cal U}(k):=\{{\bf u}(j,k), j \in {\cal J}_n\}, k \in {\cal K}_n, \]
such that
\begin{equation} \label{eq_d11a}
\mbox{ for all $k \in {\cal K}_n$ and $j \not= j' $,} [{\bf u}(j,k)={\bf x}(i), {\bf u}(j',k)={\bf x}(i')] \Rightarrow i \not= i'
\end{equation}
\begin{equation} \label{eq_d12}
|{\cal K}(i)|\ge \frac{|{\cal K}_n|| {\cal J}_n|}{2|{\cal I}_n|} \mbox{ and } |{\cal K}_0 (i,{\bf s})|\le \frac{9|{\cal J}_n||{\cal K}_n| \lambda'_n}{2|{\cal I}_n|}
\end{equation}
for all $i \in {\cal I}_n$ and ${\bf s} \in {\cal S}^n$, where
\begin{equation} \label{eq_d13}
{\cal K}(i):=\{k: \mbox{ there exists a $j \in {\cal J}_n$ with ${\bf u}(j,k)={\bf x}(i)$}\}
\end{equation}
and
\begin{equation} \label{eq_d14}
{\cal K}_0(i,{\bf s}):=\{k: \mbox{ there exists a $j$ with ${\bf u}(j,k)={\bf x}(i)$ and  $\sum_{j' \in {\cal J}_n\setminus \{j\}} tr[\rho^{\otimes n} ({\bf x}(i), {\bf s}) {\cal P}({\bf u}(j',k))] >\mu_n$}\}.
\end{equation}
Now we choose ${\cal U}(k)$ as the code book of our $k$th code $\gamma(k)$.
\subsubsection{Define Decoding Measurements}

We define its decoding measurement $\{{\cal D}(j,k), j \in {\cal J}_n\}$ for the $k$th code $\gamma (k)$, such that
\begin{equation} \label{eq_decoding}
{\cal D}(j,k):= [\sum_{j' \in {\cal J}_n} {\cal P}({\bf u}(j',k))]^{-\frac{1}{2}}{\cal P}({\bf u}(j,k)[\sum_{j' \in {\cal J}_n} {\cal P}({\bf u}(j',k))]^{-\frac{1}{2}}
\end{equation}
for its $j$th codeword ${\bf u}(j,k)$.

\subsubsection{Define the Random Correlated Code} Let our random code $\Gamma$ be randomly uniformly generated from the set of codes $\{ \gamma (k), k \in {\cal K}_n\}$.

\subsection{Error Analysis} \label{subs_err}
At first we have to estimate $tr[\rho^{\otimes n}({\bf u}(j,k), {\bf s}) {\cal P}({\bf u}(j,k))]$ for all $j \in {\cal J}_n, k \in {\cal K}_n$ and ${\bf s} \in {\cal S}^n$. To the end let us first fix $j \in {\cal J}_n, k \in {\cal K}_n$ and ${\bf s} \in {\cal S}^n$. Let $P_{XS}$ be joint type of $({\bf u}(j,k), {\bf s})$ and
\begin{equation} \label{eq_d15}
\bar{\bar{\rho}}_{S|X}(x) :=\sum_{s' \in {\cal S}}P_{S|X}(s'|x) \rho(x,s'),
\end{equation}
for all $x \in {\cal X}$. Then by (\ref{eq_doublebar}) we have that $\{\bar{\bar{\rho}}_{S|X}(x), x \in {\cal X}\} \in \bar{\bar{{\cal W}}}$. Therefore by (\ref{eq_d1}) we obtain that for ${\bf u} (j, k):=(u_1(j,k), u_2(j.k), \ldots, u_n(j,k))$
\begin{eqnarray}\label{eq_d16}
&&\sum_{{\bf s}' \in {\cal S}^n} P^n_{S|X}({\bf s}'|{\bf u} (j, k)) tr [\rho^{\otimes n} ({\bf u}(j,k), {\bf s}') {\cal P}({\bf u}(j,k))] \nonumber \\
&&=tr \{\{\sum_{{\bf s}' \in {\cal S}^n} [\prod_{t=1}^nP_{S|X}(s'_t|u_t(j,k))] [\bigotimes_{t=1}^n \rho(u_t(j,k), s'_t)]\} {\cal P}({\bf u}(j,k))\}   \nonumber \\
&&=tr \{[\bigotimes_{t=1}^n(\sum_{s'_t \in {\cal S}}P_{S|X}(s'_t|u_t(j,k))\rho(u_t(j,k), s'_t))] {\cal P}({\bf u}(j,k))\}\nonumber \\
&&=tr [\bar{\bar{\rho}}^{\otimes n}_{S|X}({\bf u}(j,k)){\cal P}({\bf u}(j,k))] > 1-2^{-n \eta},
\end{eqnarray}
where \[\bar{\bar{\rho}}^{\otimes n}_{S|X}({\bf u}(j,k))=\bar{\bar{\rho}}_{S|X}(u_1(j,k)) \otimes \bar{\bar{\rho}}_{S|X}(u_2(j,k))\otimes \ldots, \otimes \bar{\bar{\rho}}_{S|X}(u_n(j,k)).\]
However by Lemma \ref{lemma_comp} (iii), the value of $ tr [\rho^{\otimes n} ({\bf u}(j,k), {\bf s}') {\cal P}({\bf u}(j,k))]$ depends only on the joint type of $ ({\bf u}(j,k), {\bf s}')$ and so does $ tr [\rho^{\otimes n} ({\bf u}(j,k), {\bf s}') (\mathbb{I}_{\cal H}-{\cal P}({\bf u}(j,k)))]$.
Therefore (\ref{eq_d16}) yields that
\begin{eqnarray} \label{eq_d17}
&&2^{-n \eta} > \sum_{{\bf s}' \in {\cal S}^n}P^n_{S|X}({\bf s}'|{\bf u} (j, k)) tr [\rho^{\otimes n} ({\bf u}(j,k), {\bf s}') (\mathbb{I}_{\cal H}-{\cal P}({\bf u}(j,k)))]  \nonumber \\
&& \ge \sum_{{\bf s}' \in {\cal T}^n_{S|X}({\bf u}(j,k))}P^n_{S|X}({\bf s}'|{\bf u} (j, k)) tr [\rho^{\otimes n} ({\bf u}(j,k), {\bf s}') (\mathbb{I}_{\cal H}-{\cal P}({\bf u}(j,k)))]  \nonumber \\
&&=P^n_{S|X}[{\cal T}^n_{S|X}({\bf u}(j,k))|{\bf u}(j,k)]tr [\rho^{\otimes n} ({\bf u}(j,k), {\bf s}) (\mathbb{I}_{\cal H}-{\cal P}({\bf u}(j,k)))],
\end{eqnarray}
for the particular ${\bf u}(j,k)$ and ${\bf s}$, since $P_{XS}$ is the joint type of  ${\bf u}(j,k)$ and ${\bf s}$. That is,
\begin{equation}\label{eq_d18}
tr [\rho^{\otimes n} ({\bf u}(j,k), {\bf s}) (\mathbb{I}_{\cal H}-{\cal P}({\bf u}(j,k)))]< 2^{-\frac{n\eta}{2}}\end{equation}
or
\begin{equation}
tr [\rho^{\otimes n} ({\bf u}(j,k), {\bf s}) {\cal P}({\bf u}(j,k))] \ge 1- 2^{-\frac{n\eta}{2}}, \nonumber
\end{equation}
for all ${\bf u}(j,k)$ and ${\bf s}$, as $P^n_{S|X}[{\cal T}^n_{S|X}({\bf u}(j,k))|{\bf u}(j,k)]>2^{-\frac{n\eta}{2}}$ for any $\eta >0$ and sufficiently large $n$.

Let $J$ and $K$ be two independent random variables taking values in ${\cal J}_n$ and ${\cal K}_n$ according uniform distributions, respectively. Since (\ref{eq_d11a}) and (\ref{eq_d13}) yield that for very $k \in {\cal K}(i)$ there is exactly one $j:=j(i,k)$ (say) in ${\cal J}_n$, such that ${\bf u}(k,j(i,k)) ={\bf x}(i)$, by (\ref{eq_d12}) we have that for all ${\bf x} (i) \in {\cal B}, {\bf s} \in {\cal S}^n$
$Pr[{\bf u}(J, K)={\bf x}(i)]=\frac{|{\cal K}(i)|}{|{\cal J}_n||{\cal K}_n|}>0$ for all $i \in {\cal I}_n$ and
\begin{equation}\label{eq_cexp}
\mathbb{E}\{tr[\rho^{\otimes n}({\bf u}(J,K), {\bf s}) {\cal D}(J, K)]|{\bf u}(J, K)={\bf x}(i)\}=\frac{1}{|{\cal K}(i)|} \sum_{k \in {\cal K}(i)}tr[\rho^{\otimes n}({\bf x}(i),{\bf s}){\cal D}((j(i,k)),k)].\end{equation}
Next we shall apply
Hayashi-Nagaoka inequality
\begin{equation} \label{eq_HH}
\mathbb{I}_{\cal H}-(S + T )]^{-\frac{1}{2}} S (S + T )^{-\frac{1}{2}} \le 2 (\mathbb{I}_{\cal H}-S ) + 4T
\end{equation}
for any positive operators $S$ and $T$ with $0 \le S \le \mathbb{I}_{\cal H}$ and $T \ge 0$, to estimate
\[\max_{{\bf s} \in {\cal S}^n}\mathbb{E}\{tr[\rho^{\otimes n}({\bf u}(J,K), {\bf s})(\mathbb{I}_{\cal H}- {\cal D}(J, K))]|{\bf u}(J, K)={\bf x}(i)\}.\]
To this end let ${\cal K}_1(i, {\bf s}):={\cal K}(i) \setminus  {\cal K}_0(i, {\bf s})$ for all $i$ and ${\bf s}$. Then it follows from (\ref{eq_d12}) that
\begin{equation} \label{eq_d19}
\frac{|{\cal K}_0(i, {\bf s})|}{|{\cal K}(i)|}\le 9 \lambda'_n \mbox{ and } \frac{|{\cal K}_1(i, {\bf s})|}{|{\cal K}(i)|}\ge 1- 9 \lambda'_n.
\end{equation}
Consequently we have
\begin{eqnarray} \label{eq_d20}
&&\frac{1}{|{\cal K}(i)|}\sum_{k \in {\cal K}_0(i, {\bf s})}tr[\rho^{\otimes n}({\bf x}(i),{\bf s})(\mathbb{I}_{\cal H}-{\cal D}((j(i,k)),k))]  \nonumber \\
&&\le \frac{|{\cal K}_0(i, {\bf s})|}{|{\cal K}(i)|}\le 9 \lambda'_n,
\end{eqnarray}
for all ${\bf x} \in {\cal B}$ and ${\bf s} \in {\cal S}^n$. On the other hand, by (\ref{eq_d14}), (\ref{eq_decoding}), (\ref{eq_d18}), (\ref{eq_HH}) and the definitions of ${\cal K}_1(i, {\bf s})$ and $j(i,k)$, we obtain that
\begin{eqnarray} \label{eq_d21}
&&\frac{1}{|{\cal K}(i)|}\sum_{k \in {\cal K}_1(i, {\bf s})}tr[\rho^{\otimes n}({\bf x}(i),{\bf s})(\mathbb{I}_{\cal H}-{\cal D}((j(i,k)),k))]  \nonumber \\
&& \le \frac{1}{|{\cal K}(i)|}\sum_{k \in {\cal K}_1(i, {\bf s})}\{2 tr[\rho^{\otimes n}({\bf u}(j(i,k)),{\bf s})(\mathbb{I}_{\cal H}-{\cal P}({\bf u}(j(i,k))))]+ 4tr[\rho^{\otimes n}({\bf x}(i),{\bf s})\sum_{j'\in {\cal J}_n \setminus \{j(i,k)\}}{\cal P}({\bf u}(j',k))]\}  \nonumber \\
&& < \frac{1}{|{\cal K}(i)|}\sum_{k \in {\cal K}_1(i, {\bf s})}\{2^{-\frac{n\eta}{2}+1}+4 \mu_n\} \le 2^{-\frac{n\eta}{2}+1}+4 \mu_n,
\end{eqnarray}
where to have the first inequality, we first apply (\ref{eq_decoding}) and (\ref{eq_HH}) to break $(\mathbb{I}_{\cal H}-{\cal D}((j(i,k)),k))$ to two terms and then by the definition of $j(i,k)$ substitute
$\rho^{\otimes n}({\bf x}(i),{\bf s})$ by $\rho^{\otimes n}({\bf u}(j(i,k)),{\bf s})$ in the first term; the second inequality holds by (\ref{eq_d18}), (\ref{eq_d14}) and the facts that ${\cal K}_1(i, {\bf s}):={\cal K}(i) \setminus  {\cal K}_0(i, {\bf s})$ and $\rho^{\otimes n}({\bf x}(i),{\bf s})=\rho^{\otimes n}({\bf u}(j(i,k)),{\bf s})$; and finally the last inequality follows from that $|{\cal K}_1(i,{\bf s})|\le |{\cal K}(i)|$. Now (\ref{eq_cexp}), (\ref{eq_d20}) and (\ref{eq_d21}) together yield that
\begin{eqnarray} \label{eq_d22}
 &&\mathbb{E}\{tr[\rho^{\otimes n}({\bf u}(J,K), {\bf s}) (\mathbb{I}_{\cal H}-{\cal D}(J, K))]|{\bf u}(J, K)
={\bf x}(i)\} \nonumber \\
&&=\frac{1}{|{\cal K}(i)|}\sum_{k \in {\cal K}_0(i, {\bf s})}tr[\rho^{\otimes n}({\bf x}(i),{\bf s})(\mathbb{I}_{\cal H}-{\cal D}((j(i,k)),k))]\nonumber \\
&&+\frac{1}{|{\cal K}(i)|}\sum_{k \in {\cal K}_1(i, {\bf s})}tr[\rho^{\otimes n}({\bf x}(i),{\bf s})(\mathbb{I}_{\cal H}-{\cal D}((j(i,k)),k))]  \nonumber \\
&&<9 \lambda'_n+2^{-\frac{n\eta}{2}+1}+4 \mu_n,
\end{eqnarray}
for all ${\bf x}(i) \in {\cal B}$ and all ${\bf s} \in {\cal S}^n$. That is,
\begin{equation} \label{eq_d23}
\max_{{\bf s} \in {\cal S}^n}\mathbb{E}\{tr[\rho^{\otimes n}({\bf u}(J,K), {\bf s})(\mathbb{I}_{\cal H}- {\cal D}(J, K))]|{\bf u}(J, K)={\bf x}(i)\} <9 \lambda'_n+2^{-\frac{n\eta}{2}+1}+4 \mu_n,
\end{equation}
for all ${\bf x}(i) \in {\cal B}$, or
\begin{equation} \label{eq_d23a}
\sum_{{\bf x}(i)} Pr\{{\bf u}(J,K)={\bf x}(i)\}\max_{{\bf s} \in {\cal S}^n}\mathbb{E}[tr[\rho^{\otimes n}({\bf u}(J,K), {\bf s})(\mathbb{I}_{\cal H}- {\cal D}(J, K))]|{\bf u}(J, K)={\bf x}(i)] <9 \lambda'_n+2^{-\frac{n\eta}{2}+1}+4 \mu_n
\end{equation}
Consequently, by (\ref{eq_f2a}), we conclude that
\begin{equation} \label{eq_d23b}
p_a(\Gamma)<9 \lambda'_n+2^{-\frac{n\eta}{2}+1}+4 \mu_n.
\end{equation}

Finally we notice that like in the standard way to apply random choice for showing direct coding theorem in classical and quantum Shannon Theory, we have not excluded the case that for $i \not= i'$ in ${\cal I}_n$, ${\bf x}(i)$ and ${\bf x}(i')$ take the same input codeword as their values, formally distinguish them by their indices, and consider them as different members of ${\cal B}$ even in the case that it occurs. (It is the reason why we do not write ``${\bf u}(j,k) \not={\bf u}(j',k)$ for $j \not= j'$" in (\ref{eq_d11a}).) This slightly makes a difference in (\ref{eq_d23}) and (\ref{eq_d23a}). That is, if ${\bf x}(i)={\bf x}(i')={\bf x}$ and ${\bf x}$ is sent, by our assumption jammer only knows the input codeword ${\bf x}$, but does not know which index in ${\cal B}$ leads to the input codeword. On the other hand the expressions at left hand sides of (\ref{eq_d23}) and (\ref{eq_d23a}) mean that jammer may choose state sequence according to the index, which implies the jammer has more information than our assumption. Thus, in this case left hand side of (\ref{eq_d23a}) in fact is an upper bound of conditional expectation at right hand side of  (\ref{eq_f2a}). Clearly this does not impede us to have (\ref{eq_d23b}).

\subsection{Set up the Parameters} \label{subs_par}
Now we have to fix the parameters $A_n, |{\cal I}_n|, |{\cal J}_n|, |{\cal K}_n|, \mu_n$ and $\lambda'_n$ and they must 
satisfy our previous assumptions (\ref{eq_A}), (\ref{eq_I}), (\ref{eq_J1}), (\ref{eq_AJ}), (\ref{eq_JK}) and (\ref{eq_Ia}). 
Given $\epsilon>0$ (independent of $n$) and $\lambda_n$ with $\lambda_n \ge \max(2^{-\frac{n\eta}{3}}, 2^{-\frac{n\epsilon}{5}})$ (for $\eta$ in (\ref{eq_d1})), 
(which may or may not depend on $n$,) we hope to have a code with rate $\frac{1}{n} \log |{\cal J}_n| >\min_{ \bar{\bar{\rho}}(\cdot) \in \bar{\bar{\cal W}}} 
\chi(P_X, \bar{\bar{\rho}}(\cdot))-\epsilon$ and probability of error smaller than $\lambda_n$ to minimize {\it the order}, of the size of random code $|{\cal K}_n|$.

At first we note that $\xi$ and $\nu$ in (\ref{eq_d3}) can be arbitrary positive numbers, then we choose them such that $
0 <\xi+\nu< \frac{\epsilon}{2}$. Let $A_n=2^{-n[\min_{ \bar{\bar{\rho}}(\cdot) \in \bar{\bar{\cal W}}} \chi(P_X, \bar{\bar{\rho}}(\cdot))-\frac{\epsilon}{2}]}$ 
and then (\ref{eq_A}) holds. Next we choose $a_1$ as positive real larger than $\frac{1}{3-e}$ such that $a_1 \frac{n \log_e|{\cal X}|{\cal S|}}{A_n}$ is a 
integer and let $|{\cal I}_n|= a_1 \frac{n \log_e|{\cal X}||{\cal S}|}{A_n}$. Thus (\ref{eq_I})  and (\ref{eq_Ia}) hold. 
Let $a_2=\frac{1}{27}, \mu_n=\lambda'_n=a_2 \lambda_n$ so that the upper bound to the average probability of error at the right hand 
side of (\ref{eq_d23b}) is smaller than $\lambda_n$ when $n$ is sufficiently large. Let $|{\cal J}_n|=\frac{(a_2\lambda_n)^2}{A_n}=\frac{\lambda'_n \mu_n}{A_n}$ (or its integer part) and then (\ref{eq_J1}) and (\ref{eq_AJ}) hold, and $\frac{1}{n} \log |{\cal J}_n| >\min_{ \bar{\bar{\rho}}(\cdot) \in \bar{\bar{\cal W}}} \chi(P_X, \bar{\bar{\rho}}(\cdot))-\epsilon$ (since by our assumption $(a_2\lambda_n)^2 > 2^{-\frac{n\epsilon}{2}}$) . Finally to satisfy (\ref{eq_JK}), we choose
\begin{equation} \label{eq_d24}
|{\cal K}_n|=\frac{32n|{\cal I}_n|\log_e|{\cal X}||{\cal S}|}{\lambda'_n|{\cal J}_n|}=\frac{32 a_1 (n \log_e |{\cal X}||{\cal S}|)^2}{(a_2\lambda_n)^3}= \frac{an^2}{\lambda_n^3},
\end{equation}
(or its integer part) for a constant $a:=\frac{32 a_1 ( \log_e |{\cal X}||{\cal S}|)^2}{a_2^3}$ depending only on $|{\cal X}||{\cal S}|$, where the second equality is obtained by substitute $|{\cal I}_n|= \frac{a_1 n \log_e|{\cal X}||{\cal S}|}{A_n},\lambda'_n=a_2 \lambda_n$ and $|{\cal J}_n|=\frac{(a_2\lambda_n)^2}{A_n}$. Thus the proof is completed.

\section{proof of Theorem \ref{thmc}}\label{itswpt}

 In this section we prove Theorem \ref{thmc}. 
At first, we show  
Theorem \ref{thmc} for codes with vanishing key rate as those in Theorem \ref{thm_d} , i.e., when there is a positive constant $B$
such that
$|{\cal K}|\leq bn^2$.

Suppose that we are given a random correlated code $\Gamma$ taking value on $\{(\{{\bf u}(j,k), j \in {\cal J}\}, \{{\cal D}(j, k),  j \in {\cal J}\}), k \in {\cal K}\}$ such that the random message $J$ is randomly uniformly distributed on ${\cal J}$ and the random key $K$ is randomly distributed on ${\cal K}$ with any distribution. Denote the rate and the average probability of error of the code $\Gamma$ by $R$ and $\lambda$ respectively.

As a randomizing or so-called mixed strategy may not enlarge the probability of error, without loss of generality we assume the 
jammer randomly chooses state sequences, according to the input codeword. More specifically let ${\bf X}'={\bf u}(J,K)$ be the
 random input of the AVCQC and $P_{X'}$ be its distribution. Then the jammer knows both the input distribution 
$P_{{\bf X}'}$ and the outcome ${\bf x}$ of ${\bf X}'={\bf u}(J,K)$, since we assume he knows that both coding scheme and 
input codeword. Let $P_{X_t'}$ be the $t$th marginal distribution of $P_{X'}$.

Let 
\begin{equation} \label{eq_c1}
\bar{\bar{\rho}}(x):=\sum_s Q(s|x) \rho(x,s) \in \bar{\bar{\cal W}}
\end{equation}
be an arbitrary classical quantum channel in $\bar{\bar{\cal W}}$
component wise independently. That is, \[Pr\{{\bf S}={\bf s}|{\bf X}'={\bf x}\}=\prod_{t=1}^n Q_t(s_t|x_t),\]
where $Q_t$ is  the $t$th marginal distribution of $Q$.
Let $R$ be a $\epsilon$-achievable rate for $\{\bar{\bar{\rho}}_t(x):t,x\}$ with a $\epsilon\in  [0,1)$,
where $\bar{\bar{\rho}}_t(x):=\sum_{s_t} Q(s_t|x_t) \rho(x_t,s_t)$.
By Winter’s strong converse for
the single memoryless  classical quantum channel  in \cite{Win} 
for any $\delta$ when $n$ is sufficiently large
it holds
\[nR\leq \frac{1}{\left| {\cal K}\right|}\sum_{k\in  {\cal K}}\chi\left(P_{X'};\bar{\bar{\rho}}^{\otimes n} (x(\cdot,k))\right)+n\delta\]
Let ${\bf X}$ be the random variable taking value on $u({\cal J})$
such that $P_{X}^n(x(j))$  $=\sum{k\in {\cal K}} P_{X'}(x(j,k))$.
Let $G_{uni}$ be the uniformly distributed random variable with value in
${\cal K}$.
When $n$ is sufficiently large we have
\begin{align}&\frac{1}{\left| {\cal K}\right|}\sum_{k\in  {\cal K}}\chi\left(P_{X'};\bar{\bar{\rho}}^{\otimes n} (x(\cdot,k))\right)
- \chi\left(P_{X};\bar{\bar{\rho}}^{\otimes n} (x(\cdot))\right)\notag\\
&= \frac{1}{\left|{\cal K}\right|}\sum_{k\in {\cal K}}S\left(\frac{1}{|{\cal J}|}\sum_{j=1}^{|{\cal J}|}x(j,k)\right)
-\frac{1}{\left|{\cal K}\right|}\frac{1}{|{\cal J}|}\sum_{k\in {\cal K}}\sum_{j=1}^{|{\cal J}|}S\left(x(j,k)\right)\notag\\
&-S\left(\frac{1}{\left|{\cal K}\right|}\frac{1}{|{\cal J}|}\sum_{k\in {\cal K}}\sum_{j=1}^{|{\cal J}|}x(j,k)\right)
+ \frac{1}{|{\cal J}|}\sum_{j=1}^{|{\cal J}|}S\left(\frac{1}{\left|{\cal K}\right|}\sum_{k\in {\cal K}}x(j,k)\right)\allowdisplaybreaks\notag\\
&=\frac{1}{|{\cal J}|}\sum_{j=1}^{|{\cal J}|}\chi\left(G_{uni},\bar{\bar{\rho}}^{\otimes n}(x(j,k))\right)
-\chi\left(G_{uni},\sum_{j=1}^{|{\cal J}|}P_{X}(j)\bar{\bar{\rho}}^{\otimes n}(x(j,k))\right)\allowdisplaybreaks\notag\\
&\leq \frac{1}{|{\cal J}|}\sum_{j=1}^{|{\cal J}|}\chi\left(G_{uni},\bar{\bar{\rho}}^{\otimes n}(x(j,k))\right)\leq \frac{1}{|{\cal J}|}\sum_{j=1}^{|{\cal J}|} H\left(G_{uni}\right)\allowdisplaybreaks\notag\\
&=2 \log n + \log b\le n\delta \text{ .}\label{slgutbqnrn}\end{align} \vspace{0.2cm}

Now we assume that the jammer chooses the $t$th component $s_t$ of random state sequence 
${\bf S}$ according to the $t$th outcome of the random input ${\bf X}'$ and the conditional distribution $Q_t$ for $t=1,2, \ldots, n$.
By applying first Holevo bound to the ensemble $\{(P_{{\bf X}}({\bf x}),\bar{\bar{\rho}}^{\otimes n *}({\bf x})), {\bf x} \in {\cal X}^n\}$, for the classical quantum channel
\begin{equation} \label{eq_c3}\bar{\bar{\rho}}^{\otimes n *}({\bf x})=\bigotimes_{t=1}^n [\sum_{s_t} Q_t(s_t|x_t)\rho(x_t,s_t)]=\bigotimes_{t=1}^n\bar{\bar{\rho}}_t(x_t)\end{equation}
for ${\bf x}=(x_1,x_2, \ldots, x_n)$ and ${\bf s}=(s_1,s_2, \ldots, x_n)$,
and then subadditivity of von Neumann entropy we obtain that
\begin{eqnarray} \label{eq_c4}
&&nR\le \chi (P_{{\bf X}},\bar{\bar{\rho}}^{\otimes n *}(\cdot))+n\delta (\lambda)=S (\sum_{{\bf x}}P_{{\bf X}}({\bf x})
\bar{\bar{\rho}}^{\otimes n *}({\bf x}))-\sum_{{\bf x}}P_{{\bf X}}({\bf x})S (\bar{\bar{\rho}}^{\otimes n *}({\bf x})) +2n\delta\nonumber \\
&&\le \sum_{t=1}^n S(\sum_{x_t} P_{X_t}(x_t)\sum_{s_t} Q_t(s_t|x_t)\rho(x_t,s_t))-\sum_{{\bf x}}P_{{\bf X}}({\bf x})S (\bar{\bar{\rho}}^{\otimes n *}({\bf x})) +2n\delta\nonumber \\
&&= \sum_{t=1}^n S(\sum_{x_t} P_{X_t}(x_t)\sum_{s_t} Q_t(s_t|x_t)\rho(x_t,s_t))-\sum_{t=1}^n[\sum_{x_t}P_{X_t}(x_t) S(\sum_s Q_t(s_t|x_t)\rho(x_t,s_t))]+2n\delta \nonumber \\
&&= \sum_{t=1}^n [S(\sum_{x_t} P_{X_t}(x_t)\sum_{s_t} Q_t(s_t|x_t)\rho(x_t,s_t))-\sum_{x_t}P_{X_t}(x_t) S(\sum_s Q_t(s_t|x_t)\rho(x_t,s_t))]+2n\delta\nonumber \\
&&=\sum_{t=1}^n[S(\sum_{x_t}  P_{X_t}(x_t)\bar{\bar{\rho}}_t(x_t))-\sum_{x_t}  P_{X_t}(x_t)S(\bar{\bar{\rho}}_t(x_t))]+2n\delta= \sum_{t=1}^n \chi (P_{X_t}, \bar{\bar{\rho}}_t(\cdot))+2n\delta
\end{eqnarray}
where the first and the last equalities follow from the definition of Holevo quantity; 
the first inequality holds by (\ref{eq_c3}) and the subadditivity of von Neumann entropy; and the second equality follows 
from (\ref{eq_c3}); the second last equality follows from (\ref{eq_c1}).

$ \bar{\bar{\cal W}}$ is a
compact set, and 
$\chi(\cdot,\cdot)$ is a concave-convex function, therefore by the Minimax Theorem we have
\[ \max_P\min_{\bar{\bar{\rho}}(\cdot)} \chi(P, \bar{\bar{\rho}} (\cdot)) = \min_{\bar{\bar{\rho}}(\cdot)} \max_P\chi(P, \bar{\bar{\rho}} (\cdot)) . \]

From (\ref{eq_c4}) and (\ref{slgutbqnrn})  we have that
\begin{align}&R\le \min_{\bar{\bar{\rho}}(\cdot)}\frac{1}{n}\sum_{t=1}^n  \chi(P_{X_t}, \bar{\bar{\rho}} (\cdot))+ 2n\delta\nonumber \\
&\le
\min_{\bar{\bar{\rho}}(\cdot)}  \max_{P_X}\chi(P_{P_X}, \bar{\bar{\rho}} (\cdot))+2n\delta\nonumber \\
&= \max_{P_X}
\min_{\bar{\bar{\rho}}(\cdot)} \chi(P_{P_X}, \bar{\bar{\rho}} (\cdot))+2n\delta.\label{rlmbb}\end{align}
(\ref{rlmbb}) proves Theorem \ref{thmc} for codes with a vanishing key rate.\vspace{0.2cm}

Now we want to prove Theorem \ref{thmc} for codes with an arbitrary key rate
For the proof of  (\ref{slgutbqnrn}) we assume that  the
 key rate is vanishing. In fact   (\ref{slgutbqnrn})
also holds with  arbitrary key size $|{\cal K}|$ when
we  limit the amount of common randomness.
similar to the results for  classical arbitrarily varying wiretap  channel
in \cite{Wi/No/Bo}.

\begin{lemma}[cf. \cite{Bo/Ca/De2}] Let $c > 0$.
 For every $q \in P({\cal S})$
 and $s^n\in {\cal S}^n$, let a function $I_{q,s^n} : \Gamma \rightarrow
(0,c)$ be given. Assume
these functions satisfy the following:
for every $\gamma\in\Gamma$,
  $s^n\in \theta^n$, and $q,q' \in P(\theta)$ satisfy $\|q - q'\|_1 \leq \delta$
\[|I_{q,s^n} (\gamma) - I_{q',s^n}(\gamma)| \leq f(\delta)\text{  ,}\]
 for some $f(\delta)$
 which tends to $0$ as $\delta$ tends to $0$. We write $\mu(I_{q,s^n} ):=\sum_{\gamma\in\Gamma}\mu(\gamma)I_{q,s^n} (\gamma)$,
where $\mu(\gamma)$ is the probability of $\gamma$.
 Then for every $\varepsilon > 0$ and sufficiently large $n$, there are $L = n^2$
realizations $\gamma_1,\cdots, \gamma_L$ such that
\[\frac{1}{L}\sum_{l=1}^{L}I_{q,s^n} (\gamma_l)\geq(1-\varepsilon)\mu(I_{q,s^n} )-\varepsilon\]
for every $q \in P(\theta)$
 and $s^n\in \theta^n$.
\label{lcfeqicsit}\end{lemma}

For  a conditional distribution $Q$ on $\cal S$ and $\bar{\bar{\rho}}(x)$ $=\sum_s Q(\cdot|x) \rho(x,s)$
we define
\[I_{Q,s^n}(k) := \frac{1}{n}\chi(P_{X};\chi(P_{X};\bar{\bar{\rho}}^{\otimes n}(x(j,k)))\text{  .}\]

In \cite{Bo/No2} the continuity of 
$ Q(\cdot|x) \rightarrow \sum_s Q(\cdot|x) \rho(x,s)$
 has been shown; thus 
 when for any conditional distribution $Q'$ on $\cal S$ fulfilling
$\|Q(~|x) - Q'(~|x)\|_1 =\delta\rightarrow 0$ for all $x$ 
there is a 
$f(\delta)$ such that $|I_{Q,s^n} (k) - I_{Q',s^n}(k)|$
$\frac{1}{n}\frac{1}{\left|{\cal K}\right|} \sum_{k=1}^{\left|{\cal K}\right|}\chi(P_{X};\bar{\bar{\rho}}^{\otimes n}(x(j,k))$ $-$
$\frac{1}{n}\frac{1}{\left|{\cal K}\right|} \sum_{k=1}^{\left|{\cal K}\right|}(\chi(P_{X};{\bar{\bar{\rho}}^{\otimes n}}'(x(j,k)))$
$\leq f(\delta)$ for a  $f(\delta)$ that fulfills $f(\delta) \rightarrow 0$, where ${\bar{\bar{\rho}}'}(x)$ 
$:=  \sum_s Q'(s|x) \rho(x,s)$.
By Lemma \ref{lcfeqicsit}
there is a set ${\cal K}'\subset{\cal K}$ such that $\left|{\cal K}'\right| = n^2$ and
\begin{align*}&
\frac{1}{\left|{\cal K}'\right|}\frac{1}{n}\sum_{k'\in {\cal K}'}
\chi(P_{X}\bar{\bar{\rho}}^{\otimes n}(x(j,k')))\\
&\geq (1-\varepsilon)\frac{1}{n}
\frac{1}{\left|{\cal K}\right|}\sum_{k\in {\cal K}}
\chi(P_{X};\bar{\bar{\rho}}^{\otimes n}(x(j,k)))  \text{  .}\end{align*}

Thus
\begin{align}& \frac{1}{n}\log |{\cal J}|\notag\\
&\leq \frac{1}{1-\varepsilon}
\frac{1}{n}\frac{1}{\left|{\cal K}'\right|}\sum_{k\in {\cal K}'}
\left(\chi(P_{X};\bar{\bar{\rho}}^{\otimes n}(x(j,k)))+\delta\right)\notag\\
&\leq  \frac{1}{1-\varepsilon}\frac{1}{n}
 \max_{P_X}
\min_{\bar{\bar{\rho}}(\cdot)} \chi(P_{P_X}, \bar{\bar{\rho}} (\cdot))+2\delta
\text{  .}\label{1nlcjlf}\end{align}

(\ref{1nlcjlf}) shows that
(\ref{slgutbqnrn}) 
is even then true if we
do not have a vanishing key rate, i.e., when we do not have $|{\cal K}|\leq bn^2$.

\section{proof of Theorem \ref{thm_da}}\label{sec_prda}

The proof will be done by modification of the step 3 of proof of Theorem \ref{thm_d} in Subsection \ref{subs_con} of Section \ref{sec_prd} to have a  code achieving the full capacity
not only in scenario 1,
but also in scenario 2, as follows.

Let a ground set of codeword ${\cal B}=\{{\bf x}(i), i \in {\cal I}\}$ be generated in Subsection \ref{subs_gr} and $A_n, |{\cal I}_n|, |{\cal J}_n|, |{\cal K}_n|, \mu_n$ and $\lambda'_n$ be given in Subsection \ref{subs_par}. Additionally, without loss of generality, we require $|{\cal I}_n|$ is divided by $|{\cal J}_n|$, i. e.,$B_n:=\frac{|{\cal I}_n|}{|{\cal J}_n|}=\frac{na_1\log_e |{\cal X}||{\cal S}|}{(a_2\lambda_n)^2}$ is an integer. Thus we may partition $|{\cal I}_n|$ into $|{\cal J}_n|$ subsets, ${\cal I}_n(j), j \in {\cal J}_n$ with equal size $B_n=\frac{|{\cal I}_n|}{|{\cal J}_n|}$ in an arbitrary way. Let ${\cal B}(j)=\{{\bf x}(i): i \in {\cal I}_n(j)\}$ for $j\in {\cal J}_n$. Let ${\bf U}'(j,k)$ be independently and uniformly generated from ${\cal B}(j)$ for $j \in {\cal J}_n$ respectively and all $k \in {\cal K}_n$. Then for all ${\bf x} \in {\cal T}^n_X, {\bf s} \in {\cal S}^n$ and $k \in {\cal K}_n$, we have that
\begin{eqnarray} \label{eq_da1}
&&\mathbb{E}tr[\rho^{\otimes n}({\bf x},{\bf s}) \sum_{j \in {\cal J}_n} {\cal P}({\bf U}'(j,k))] =\sum_{j \in {\cal J}_n}\mathbb{E}tr[\rho^{\otimes n}({\bf x},{\bf s}) {\cal P}({\bf U}'(j,k))] \nonumber \\
&&=\sum_{j \in {\cal J}_n}[\sum_{i(j) \in {\cal I}_n(j)}\frac{1}{B_n}tr[\rho^{\otimes n}({\bf x},{\bf s}){\cal P}({\bf x}(i(j)))]\nonumber \\
&&=\frac{1}{B_n}\sum_{i \in {\cal I}_n}tr[\rho^{\otimes n}({\bf x},{\bf s}){\cal P}({\bf x}(i))]\le 3A_n|{\cal J}_n|,
\end{eqnarray}
where the last equality holds because $\{{\cal I}_n(j), j \in {\cal J}_n\}$ is a partition of ${\cal I}_n$; and the last inequality follows from (\ref{eq_d6}) and $B_n=\frac{|{\cal I}_n|}{|{\cal J}_n|}$. Because of the independence of ${\bf U}'(j,k), j \in {\cal J}_n$, by Markov inequality we have that for all $j \in {\cal J}_n, i(j) \in {\cal I}_n(j)$ and ${\bf s} \in {\cal S}^n$,
\begin{eqnarray} \label{eq_mara}
&&Pr\{\sum_{j' \in {\cal J}_n\setminus \{j\}} tr[\rho^{\otimes n} ({\bf U}'(j,k), {\bf s}) {\cal P}({\bf U}'(j',k))] >\mu_n |{\bf U}'(j,k)={\bf x}(i(j))\}]\nonumber \\
&&\le \frac{\mathbb{E}\{\sum_{j' \in {\cal J}_n\setminus \{j\}} tr[\rho^{\otimes n} ({\bf x}(i(j)), {\bf s}) {\cal P}({\bf U}'(j',k))] |{\bf U}'(j,k)={\bf x}(i(j))\}}{\mu_n}\nonumber \\
&&\le \frac{\mathbb{E}\{\sum_{j' \in {\cal J}_n} tr[\rho^{\otimes n} ({\bf x}(i(j))), {\bf s}) {\cal P}({\bf U}'(j',k))] \}}{\mu_n}\nonumber \\
&& \le \frac{3A_n|{\cal J}_n|}{\mu_n},
\end{eqnarray}
which is analogue to (\ref{eq_mar}), where the fist inequality is Markov inequality; the second inequality holds because ${\bf U}'(j,k), j \in {\cal J}_n$ are independent and each with probability one not small than $0$; the last inequality follows from (\ref{eq_da1}).

Next for all $j \in {\cal J}_n, i(j) \in {\cal I}_n(j), k \in {\cal K}_n$ and ${\bf s} \in {\cal S}^n$, let ${\cal E}'(i(j),{\bf s},k;\mu_n)$ be the random event that
${\bf U}'(j,k)={\bf x}(i(j))$ and
\[\sum_{j' \in {\cal J}_n\setminus \{j\}} tr[\rho^{\otimes n} ({\bf x}(i(j)), {\bf s}) {\cal P}({\bf U}'(j',k))] >\mu_n,\] and
\[Z'_{i(j)}(k)=\left\{\begin{array}{ll} 1 & \mbox{if ${\bf U}'(j,k) ={\bf x}(i(j))$} \\
                                     0 & \mbox{else.} \end{array}
                                     \right. \]
Then we have that for all $j \in {\cal J}_n, i(j) \in {\cal I}_n(j)$ and $k \in {\cal K}_n$
\begin{equation} \label{eq_da2}
Pr\{Z'_{i(j)}(k)=1\}=\frac{1}{B_n}=\frac{|{\cal I}_n|}{|{\cal J}_n|},
\end{equation}
and analogously to (\ref{eq_d8})
\begin{eqnarray} \label{eq_da3}
&&Pr\{{\cal E}'(i(i),{\bf s},k;\mu_n)\}\nonumber \\
&&= Pr ({\bf U}'(j,k)={\bf x}(i(j)))Pr\{\sum_{j' \in {\cal J}_n\setminus \{j\}} tr[\rho^{\otimes n} ({\bf x}(i(j)), {\bf s}) {\cal P}({\bf U}'(j',k))] >\mu_n|{\bf U}'(j,k)={\bf x}(i(j))\} \nonumber \\
&&< \frac{3A_n|{\cal J}_n|}{B_n\mu_n}=\frac{3A_n|{\cal J}_n|^2}{|{\cal I}_n|\mu_n}.
\end{eqnarray}
Thus as we did in Subsection \ref{subs_con}, by Lemma \ref{lemma_chernoff}, ${\bf U}'(j,k), j \in {\cal J}_n, k \in {\cal K}_n$ has a realization ${\bf u}'(j,k), j \in {\cal J}_n, k \in {\cal K}_n$ with
\begin{equation} \label{eq_da4}
{\bf u}'(j,k) \in {\cal B}(j)
\end{equation}
for all $j \in {\cal J}_n$ and $k \in {\cal K}_n$ (which implies that $i \not= i'$ if ${\bf u}'(j,k)={\bf x}(i)$ and ${\bf u}'(j',k)={\bf x}(i')$ for $j \not= j'$),
\[
|{\cal K}'(i(j))|\ge \frac{|{\cal K}_n|| {\cal J}_n|}{2|{\cal I}_n|} \mbox{ and } |{\cal K}'_0 (i(i),{\bf s})|\le \frac{9|{\cal J}_n||{\cal K}_n| \lambda'_n}{2|{\cal I}_n|}
\]
for
\[
{\cal K}'(i(j)):=\{k: {\bf u}(j,k)={\bf x}(i(j))\}
\]
and
\[
{\cal K}'_0(i(j),{\bf s}):=\{k: {\bf u}(j,k)={\bf x}(i(j)) \mbox{ and }\sum_{j' \in {\cal J}_n\setminus \{j\}} tr[\rho^{\otimes n} ({\bf x}(i(j)), {\bf s}) {\cal P}({\bf u}(j',k))] >\mu_n\}.
\]
Then it follows the rest part of proof of Theorem \ref{thm_d} in Section \ref{sec_prd},
 we obtain a RCWJKI code with rate $\min_{ \bar{\bar{\rho}}(\cdot) \in \bar{\bar{\cal W}}} \chi(P_X, \bar{\bar{\rho}}(\cdot))-\epsilon$, 
average probability of error $\lambda_n$ and size $\frac{an^2}{\lambda^3}$. 
Now the   scenario 1 here, for which we have now constructed
 a code,   is actually  scenario 2, too, because by (\ref{eq_da4}), 
that the jammer knows the input codeword ${\bf u}'(j,k)$ implies that he knows the message $j$ as well. Thus our proof is completed.
\appendices
\section{proof of Lemma \ref{lemma_chernoff}} \label{sec_app}
Now let us show Lemma \ref{lemma_chernoff}
\[\begin{array}{lllllllll}
&&Pr\{\sum_{l=1}^LB_j > Lp_1(1+\alpha)\}    \\
&=& Pr\{exp_e[-\frac{\alpha}{2}Lp_1(1+\alpha)+ \frac{\alpha}{2}
\sum_{j=1}^LB_l ]>1\}
\\
&\le&exp_e[-\frac{\alpha}{2}Lp_1(1+\alpha)] \prod_{l=1}^L \mathbb{E}e^{\frac{\alpha}{2}B_l} \\
&=&exp_e[-\frac{\alpha}{2}Lp_1(1+\alpha)]
\prod_{l=1}^L[(1-p)+e^{\frac{\alpha}{2}}p]    \\
&\le&\exp_e[-\frac{\alpha}{2}Lp_1(1+\alpha)][1+(e^{\frac{\alpha}{2}}-1)p_1]^L
\\
&<&\exp_e[-\frac{\alpha}{2}Lp_1(1+\alpha)]
[1+(\frac{\alpha}{2}+\frac{e{\alpha}^2}{8})p_1]^L   \\
&<&\exp_e\{[-\frac{\alpha}{2}Lp_1(1+\alpha)]+
(\frac{\alpha}{2}+\frac{{e\alpha}^2}{8})Lp_1\}   \\
&=&\exp_E\{-\frac{\alpha}{2}Lp_1[(1+\alpha)-(1+\frac{e\alpha}{4})]\}
\\
&<&e^{-\frac{{\alpha}^2}{8}Lp_1},  \end{array}\] where the first
inequality follows from Markov inequality and the assumption $B_1,
B_2, \ldots, B_L$ are independent; the third and fourth inequalities
follows from the inequalities $e^x<1+x+\frac{e}{2}x^2$ for $x \in
(0,1)$ and $1+x <e^x$ for $x>0$ respectively. That is
(\ref{eq_lemmach1}). Similarly instead of the inequalities
$e^x<1+x+\frac{e}{2}x^2$ for $x \in (0,1)$ and $1+x <e^x$ for $x>0$
we use $e^{-x}<1-x+\frac{1}{2}x^2$ for $x \in (0,1)$ and $1-x
<e^{-x}$ for $x>0$ and have
\[\begin{array}{lllllllll}
&&Pr\{\sum_{l=1}^LB_l < Lp_0(1-\alpha)\}    \\
&=& Pr\{exp_e[\frac{\alpha}{2}Lp_0(1-\alpha)- \frac{\alpha}{2}
\sum_{l=1}^LB_l ]>1\}
\\
&\le&exp_e[\frac{\alpha}{2}Lp_0(1-\alpha)] \prod_{l=1}^l \mathbb{E}e^{-\frac{\alpha}{2}B_l} \\
&=& exp_e[\frac{\alpha}{2}Lp_0(1-\alpha)] \prod_{l=1}^L[(1-p)
+e^{-\frac{\alpha}{2}}p]     \\
&\le&\exp_e[\frac{\alpha}{2}Lp_0(1+\alpha)][1-(1-e^{\frac{-\alpha}{2}})p_0]^L
\\
&<&\exp_e[\frac{\alpha}{2}Lp_0(1-\alpha)]
[1-(\frac{\alpha}{2}-\frac{{\alpha}^2}{8})p_0]^L   \\
&<&\exp_e\{[\frac{\alpha}{2}Lp_0(1-\alpha)]-
(\frac{\alpha}{2}-\frac{{\alpha}^2}{8})Lp_0\}   \\
&=&\exp_E\{\frac{\alpha}{2}Lp_0[(1-\alpha)-(1-\frac{\alpha}{4})]\}
\\
&<&e^{-\frac{{3\alpha}^2}{8}Lp_0}.  \end{array}\] that is
(\ref{eq_lemmach2}).

\section*{Acknowledgment}
Support by the Bundesministerium f\"ur Bildung und Forschung (BMBF)
via Grant 16KIS0118K is gratefully acknowledged.

\end{document}